\documentclass{aa}  
\usepackage{graphicx}
\usepackage{txfonts}
\usepackage{lipsum}
\usepackage{subcaption}  
\usepackage{ulem}
\usepackage{lscape}             
\usepackage{placeins}           
\usepackage{hyperref}

\begin{document}
   \title{The connection between dusty star-forming galaxies and the first massive quenched galaxies}

   \author{Pablo Araya-Araya\inst{1,2,3}\thanks{\email{paaar@dtu.dk}} \and 
Rachel K. Cochrane\inst{4} \and 
Laerte Sodré Jr.\inst{3} \and Robert M. Yates\inst{5} \and Christopher C. Hayward\inst{6} \and \\ Marcel P. van Daalen\inst{7} \and Marcelo C. Vicentin \inst{3} \and Bitten Gullberg \inst{1,2} \and Francesco Valentino \inst{1,2}}

   \institute{Cosmic Dawn Center (DAWN), Copenhagen, Denmark 
\and DTU Space, Technical University of Denmark, Elektrovej 327, DK2800 Kgs. Lyngby, Denmark
\and Departamento de Astronomia, Instituto de Astronomia, Geofísica e Ciências Atmosféricas,
Universidade de São Paulo, 
Rua do Matão 1226, Cidade Universitária, 05508-900, São Paulo, SP, Brazil
\and Jodrell Bank Centre for Astrophysics, University of Manchester, Oxford Road, Manchester M13 9PL, UK
\and Centre for Astrophysics Research, University of Hertfordshire, Hatfield, AL10 9AB, UK
\and J.P. Morgan Securities LLC, 390 Madison Avenue, New York, NY 10017, USA
\and Leiden Observatory, Leiden University, PO Box 9513, NL-2300 RA Leiden, the Netherlands}

   \date{Received September 26, 2025}

\abstract{High-redshift ($z \gtrsim 2$) massive quiescent (MQ) galaxies provide an opportunity to probe the key physical processes driving the fuelling and quenching of star formation in the early Universe. Observational evidence suggests a possible evolutionary link between MQs and dusty star-forming galaxies (DSFGs; or submillimetre galaxies), another extreme high-redshift population. However, galaxy formation models have historically struggled to reproduce these populations — especially simultaneously — limiting our understanding of their formation and connection, particularly in light of recent JWST findings. In previous work, we presented a re-calibrated version of the \texttt{L-Galaxies} semi-analytic model that provides an improved match to observationally-inferred number densities of both DSFG and MQ populations. In this work, we use this new model to investigate the progenitors of MQs at $z > 2$ and the physical mechanisms that lead to their quenching. We find that most MQs at $z > 2$ were sub-millimetre-bright ($S_{870} \gtrsim 1\,{\rm mJy}$) at some point in their cosmic past. The stellar mass of MQs is strongly correlated with the maximum submillimetre flux density attained over their history, and this relation appears to be independent of redshift. However, only a minority of high-redshift DSFGs evolve into MQs by $z =2$. The key distinction between typical DSFGs and MQ progenitors lies in their merger histories: MQ progenitors experience an early major merger that triggers a brief, intense starburst and rapid black hole growth, depleting their cold gas reservoirs. In our model, AGN feedback subsequently prevents further gas cooling, resulting in quenching. In contrast, the broader DSFG population remains sub-millimetre-bright, with star formation proceeding primarily via secular processes, becoming quenched later. These findings provide a coherent theoretical framework for the formation of high-redshift MQs and clarify their connection to DSFGs, highlighting the role of mergers and AGN feedback in shaping the evolution of the most massive galaxies in the early Universe.}


   \maketitle \nolinenumbers

\section{Introduction} \label{sec:intro}
In recent years, the identification of numerous massive quiescent galaxies (MQs; $M_{\star} \gtrsim 10^{10.5}\,\rm{M_{\odot}}$ and sSFR $\lesssim 10^{-10}\,\mathrm{yr}^{-1}$) at high-redshift ($z \gtrsim 2$) has posed significant challenges to existing theoretical models of galaxy formation and evolution. Early efforts using deep photometric data — such as those from the FourStar Galaxy Evolution (ZFOURGE) survey — suggested the presence of such galaxies at $z \gtrsim 3$, prior to the peak of the cosmic star-formation rate density and the peak of the cosmic merger rate density \citep{fontana09, spitler14, straatman14}. These initial MQ candidates were selected based on red rest-frame optical colours (indicative of Balmer breaks) derived from spectral energy distribution (SED) fitting, alongside dropout techniques (e.g. the Lyman break). Subsequent spectroscopic confirmations have enabled more robust characterisations of MQs \citep[e.g.][]{glazebrook17, schreiber18, tanaka19, deugenio20, forrest20, valentino20, kubo21, nanayakkara22}, an effort extended to fainter, redder galaxies with the advent of JWST \citep{carnall23, valentino23,carnall24,nanayakkara24,alberts24, glazebrook24,long24,park24,baker25,nanayakkara25}. These observations suggest that MQs formed through intense, short-lived starbursts (SFR $\sim 500\,\rm{M_{\odot}}\ \mathrm{yr}^{-1}$) occurring within the first $\lesssim1\,\rm{Gyr}$ of cosmic time, with star formation episodes lasting $\sim 600\,\rm{Myr}$ \citep{glazebrook17, forrest20, carnall23}. The high inferred past star formation rates suggest a potential evolutionary connection with dusty star-forming galaxies (DSFGs) at higher redshifts. Other statistical properties — such as their typical number densities \citep{valentino20} and host halo masses inferred from clustering analyses \citep[e.g.][]{Wilkinson17} — also support a connection between these populations. \\
\indent The mechanisms driving the early quenching of high-redshift MQs remain unclear. X-ray and radio emission have been detected in a subset of MQs \citep{marsan17, ito22, kubo22}. Notably, MQs tend to host stronger AGN activity — as indicated by both X-ray and radio luminosities — than star-forming galaxies of similar stellar mass, whose relative excess increases with redshift \citep{ito22}. This supports the view that AGN feedback may be the dominant quenching mechanism for MQs at $z \gtrsim 3$. On the other hand, other studies \citep[e.g.,][]{Ellison22, Ellison24, Calderon-Castillo24, Gordon26} suggest that mergers primarily deplete galaxies’ cold-gas reservoirs through intense starburst episodes, rather than through AGN-driven feedback. However, the relatively short detectability timescales of AGN \citep[$\Delta t \sim 10^5\,$yr; ][]{schawinski15} may result in the underestimation of the true prevalence of recent AGN activity in quenching galaxies. MQs have also been identified in protocluster environments at $z \sim 3$; the results of \citet{kubo21}, \citet{mcconachie22}, and \citet{ito23} support a scenario in which these galaxies quenched due to accelerated internal evolution — such as elevated star formation rates combined with AGN and/or SN feedback — rather than explicit environmental effects such as ram pressure stripping. In contrast, \citet{shi21} reported evidence for environmental quenching in the D4UD01 protocluster at $z = 3.24$, suggesting that multiple quenching pathways may coexist.\\
\indent While MQs are rare, with number densities of $\sim10^{-5}\ \mathrm{Mpc}^{-3}$ at $z \sim 3.5$ \citep{schreiber18, straatman14, merlin19} (though estimates vary significantly between studies; see the compilation presented by \citealt{valentino23}), their number densities at high redshift are in tension with predictions from most current galaxy formation models. Accurately reproducing the observed abundance of MQs at $z \gtrsim 3$ at the same time as matching low-redshift constraints has proven particularly challenging; this tension has only been exacerbated by recent JWST results, which suggest an even more abundant MQ population than was inferred pre-JWST. A recent systematic comparison by \citet{lagos25} assessed the success of various semi-analytic models (SAMs) — including \texttt{GAEA} \citep{delucia24}, \texttt{GALFORM} \citep{lacey16}, and \texttt{SHARK} v2.0 \citep{lagos24} — as well as hydrodynamical simulations such as \texttt{SIMBA}, \texttt{IllustrisTNG}, and \texttt{EAGLE}, in reproducing the number densities of MQs at $z \gtrsim 3$. \citet{lagos25} showed that most of these models under-predict the observationally-inferred MQ number densities at $z \sim 4$, by between $0.3\,\rm{dex}$ and more than an order of magnitude. Similar discrepancies have been reported by other groups — for example, \citet{szpila25} using the \texttt{SIMBA-C} simulation \citep{hough23, hough24}, and \citet{vani25} using an updated version of the \texttt{L-Galaxies} SAM \citep{ayromlou21}. On the other hand, the observed number densities of MQs at high redshift could be significantly contaminated, for example by dusty galaxies or `Little Red Dots' \citep{Yang25}, a population of compact, presumably AGN-dominated sources \citep{Labbe23, Greene24, Matthee24, Li25}, recently uncovered by JWST observations.\\
\indent While some models succeed in broadly reproducing observationally-inferred number densities of MQs, those that do typically struggle with the sub-millimetre galaxy number counts. For instance, although the \texttt{IllustrisTNG} simulation yields reasonably accurate predictions for MQ number densities \citep{lagos25}, it significantly underpredicts sub-millimetre galaxy number counts \citep{chris21}. A similar situation is seen with the \texttt{EAGLE} simulation, which also matches MQ number densities at high redshift. However, as shown by \citet{cowley19}, who modelled sub-millimetre emission using the \texttt{SKIRT} radiative transfer code, the predicted sub-millimetre counts fall more than an order of magnitude below observational estimates (see also \citealt{mcalpine19}).\\
\indent The inverse trend is found in models that successfully reproduce the sub-millimetre number counts. For example, \citet{lagos19} demonstrated that the \citet{lagos18} (v1.0) version of \texttt{SHARK} matches observed sub-millimetre number counts without requiring modification of the stellar initial mass function (IMF). However, this version underpredicts the MQ number densities at $z \gtrsim 2$ by roughly $1\,\rm{dex}$ compared to the updated v2.0 release \citep{lagos24}, which better matches the lower limits of observational estimates. A similar outcome is seen with the original \texttt{Illustris} simulation, which performs well in reproducing sub-millimetre number counts \citep{chris21} yet significantly underestimates the abundance of quiescent galaxies \citep{merlin19}. Although \texttt{SIMBA} predicts a higher number density of MQs than \texttt{Illustris}, it still falls short of matching observed abundances. At the same time, \texttt{SIMBA} achieves a good match to sub-millimetre number counts \citep{lovell21}.\\
\indent This trend — where models that reproduce MQs at high redshift tend to underpredict DSFGs, and vice versa — reveals a persistent tension in galaxy formation models. It suggests that the physical mechanisms required for efficient, dust-obscured starbursts may be at odds with those needed for rapid quenching and MQ formation. Since observational evidence points towards an evolutionary connection between DSFGs and MQs \citep{daddi10, tacconi10, toft14, casey14, valentino20, forrest20}, simulations could offer a powerful framework for exploring this link. However, due to the aforementioned inconsistencies, this connection remains poorly studied from a theoretical perspective.\\
\indent Recently, \citet{yo25a}, through the application of a robust calibration method to the \citet{henriques20} version of \texttt{L-Galaxies}, tested a range of observational constraints aimed at reconciling the simultaneous modelling of MQs and DSFGs. Among nine alternative configurations, one model was found to reproduce both populations reasonably well. In this context, the re-calibrated version of \texttt{L-Galaxies} presented by \citet{yo25a} provides a unique opportunity to explore the proposed connection between high-redshift MQs and DSFGs. In this work, we use the \citet{yo25a} model initially labelled `no HIMF'\footnote{Model calibrated without using the neutral hydrogen mass function as an observational constraint; see Section \ref{sec:sam} for a detailed description.} to investigate this observationally motivated connection. In particular, we focus on identifying the progenitors of MQs at $z = 2$, $3$, and $4$, and investigating the primary physical mechanisms responsible for transforming these progenitors into massive, quiescent galaxies by these early epochs. \\
\indent The structure of this paper is as follows. In Section~\ref{sec:sam}, we provide a brief overview of the \citet{henriques20} version of the \texttt{L-Galaxies} SAM and the re-calibrated model introduced by \citet{yo25a}. Our results are presented in Section~\ref{sec:results}. In Section~\ref{sec:discussion}, we discuss our main findings. Finally, we summarise our conclusions in Section~\ref{sec:summary}.\\
\indent In this work, we adopt the \citet{planck14} cosmology: $\sigma_8= 0.829$, $H_0 = 67.3\,\rm{km\,s}^{-1}\,\rm{Mpc}^{- 1}$, $\Omega_{\Lambda}= 0.685$, $\Omega_m = 0.315$, $\Omega_b = 0.0487$, $f_b = 0.155$, and $n = 0.96$, consistent with the cosmologically rescaled version of the \texttt{Millennium} simulation \citep{angulo15}. Throughout this study, we assume a \citet{chabrier03} IMF. 

\section{Galaxy formation model} \label{sec:sam}

We use the \citet{henriques20} version of the \texttt{L-Galaxies} semi-analytical model (SAM) of galaxy formation, specifically the re-calibrated version from \citet{yo25a}. We run the model on the \texttt{Millennium} simulation \citep{springel05}, scaled to the \citet{planck14} cosmology using the method from \citet{angulo15}. This scaled version of the \texttt{Millennium} simulation has a total box volume of $(713.6\,\rm{cMpc})^{3}$ and a dark matter particle mass resolution of $m_p = 1.43 \times 10^9\,\rm{M_{\odot}}$ ($h = 0.673$, in \citet{planck14} cosmology). In this section, we describe the key features of the galaxy formation model.\\
\indent The merger trees of dark matter halos, constructed with the \texttt{SUBFIND} algorithm \citep{springel01}, form the backbone of \texttt{L-Galaxies}. These virialised halos are populated with baryonic matter, beginning with primordial diffuse gas. Initially, the gas infalls into the halos, forming a hot atmosphere. Upon cooling, this gas settles into a disk, where it becomes available for star formation, regulated by the surface density of molecular hydrogen ($\rm{H}_2$). Over time, stars die and release energy, enriching the interstellar medium (ISM) and circumgalactic medium (CGM) with metals. The energy from supernovae (SNe) can (1) heat the cold gas in the disk, moving it to hot atmosphere, available for further cooling, and (2) drive outflows that move gas to an external reservoir, from which it may later re-accrete. The \citet{henriques20} version of the model also includes metal enrichment from AGB stars, as well as from SN-Ia and SN-II events. All these processes are tracked in concentric rings within galaxies.\\
\indent Galaxy mergers are key astrophysical events that can trigger starbursts, destroy disks, form or enhance bulges, drive the growth of supermassive black holes (SMBHs), and power AGN feedback -- a critical mechanism for regulating star formation in massive galaxies. Environmental effects, such as ram pressure stripping, tidal stripping, and tidal disruption, are also modelled when a halo is accreted by a more massive one. These processes can strip hot gas, modify morphology, or even disperse the stars from the infalling galaxy into the stellar halo. \\
\indent All of these processes are described by differential equations involving free parameters, which have been robustly calibrated using MCMC methods since the \citet{henriques13} version of \texttt{L-Galaxies}. The calibration involves testing multiple sets of the free parameters and comparing model predictions with observational constraints, such as stellar mass functions (SMFs), the fraction of quiescent galaxies ($f_{\rm Q}$), the neutral hydrogen mass function (HIMF), and the cosmic star formation rate density (CSFRD), among others. To date, \texttt{L-Galaxies} is the only SAM that systematically applies a robust calibration using a Markov Chain Monte Carlo (MCMC) algorithm, without requiring additional visual inspection. \\
\indent In \citet{yo25a}, we introduced a novel calibration framework for \texttt{L-Galaxies}, testing multiple combinations of observational constraints by varying the source of the observables and omitting some constraints. Our primary goal was to determine whether a robust calibration method could find a model for reconciling the tension of modelling dusty star-forming and massive quiescent galaxies at high redshift simultaneously. Although \citet{yo25a} highlighted the high degree of degeneracy in \texttt{L-Galaxies} — and likely in all galaxy formation models — where models with distinct underlying physics can yield similar likelihoods, we identified a set of parameters that reasonably match the sub-millimetre number counts ($S_{870}$) and are consistent with lower limits on the number density of massive quiescent galaxies at high redshift. This model, referred to as ‘no HIMF’, was calibrated using the stellar mass function (SMF) and quiescent galaxy fraction ($f_{\rm Q}$) at $z = 0.4$ and $z = 2.8$ from \citet{leja20} and \citet{leja22}, respectively, along with the number density of bright sub-millimetre galaxies ($S_{870} \geq 5.2 \, {\rm mJy}$) at $z = 2.8$ from \citet{dud20}. A list of the 15 re-calibrated free parameters is presented in Table \ref{tab:freeParams}. A detailed description of these free parameters is provided in the Supplementary Material of \citet{henriques20}. \\
\begin{table}
\large \centering
\caption{Summary of the \texttt{L-Galaxies} model used in this paper.} 
\label{tab:freeParams}
\begin{tabular}{lcc}
\hline
Parameter & Value & Units \\
\hline
$\alpha_{\rm SF}$ & $0.04$ & -- \\
$\alpha_{\rm SF, burst}$ & $0.93$ & --\\
$\beta_{\rm burst}$ & $0.15$ & -- \\
$k_{\rm AGN}$ & $0.07$ & $\rm{M_{\odot}} \, {\rm yr}^{-1}$\\
$f_{\rm BH}$ & $0.04$ & --\\
$V_{\rm BH}$ & $3.92$ & ${\rm km \, s}^{-1}$\\
$\eta_{\rm reheat}$ & $0.93$ & -- \\
$V_{\rm reheat} $ & $199$ & ${\rm km \, s}^{-1}$ \\
$\beta_{\rm reheat}$ & $0.89$ & --\\
$\eta_{\rm eject}$ & $0.02$ & -- \\
$V_{\rm eject}$ & $128$ & ${\rm km \, s}^{-1}$ \\
$\beta_{\rm eject}$ & $4.38$& --  \\
$\gamma_{\rm reinc}$ & $2.5 \times 10^{10}$ & ${\rm yr}^{-1}$ \\
$\alpha_{\rm friction}$ & $6.40$ & -- \\
$M_{\rm RP}$ & $1.7\times 10^{12}$ & $\rm{M_{\odot}}$ \\
\hline
\end{tabular}
\tablefoot{These best-fit parameters were obtained by \citet{yo25a} for the `no HIMF' model, which obtains a good match to the $S_{870}$ number counts while remaining consistent with observationally-derived
lower limits on the number density of high-redshift massive quiescent galaxies. The model was calibrated using the MCMC mode of \texttt{L-Galaxies}. The observational constraints were the stellar mass function and fraction of quiescent ($\mathrm{sSFR} \leq 10^{-11}\,\mathrm{yr}^{-1}$) galaxies as a function of stellar mass at $z=0.4$ and $z=2.8$ \citep{leja20,leja22}, and the number density of bright ($S_{870} \geq 5.2\,\mathrm{mJy}$) sub-millimetre galaxies at $z=2.8$ \citep{dud20}. Full details of these parameters and their corresponding equations are given in the Supplementary Material of \citet{henriques20}.}
\end{table}
\indent The main physical mechanisms that enable this model to provide a reasonable match to both DSFG and MQ number counts at high redshift are outlined below, and discussed in more detail in \cite{yo25a}. One key aspect of the model is the importance of merger-induced starbursts. In this model, the star-formation efficiency during mergers is high, reaching up to $\sim 90\%$ in equal-mass ($1:1$) mergers. Hence, starbursts play a key role in rapidly forming stars and depleting cold gas reservoirs. As previously mentioned, mergers are also the primary channel for the growth of supermassive black holes (SMBHs). In this model, the fraction of cold gas accreted by the SMBH during a merger is fixed at $\sim 3\%$, independent of halo mass. Interestingly, among the models tested in \citet{yo25a}, this specific implementation has the highest AGN feedback strengths and provides the best agreement with the observed SMBH mass function at $z = 0$. The efficiency of gas reheating due to SN feedback in this model lies between those that best match the sub-millimetre number counts and those that most accurately reproduce the massive quiescent population. However, the fraction of SN energy allocated to driving outflows is the lowest for intermediate- and high-mass systems, implying that less energy is used to eject gas from galaxies in these halos.\\
\indent It is important to note that the \citet{henriques20} version of \texttt{L-Galaxies} does not explicitly track dust formation and destruction — a feature recently introduced in the \citet{yates24} version — nor does it directly compute sub-millimetre fluxes. Therefore, we adopt the same approach as in \citet{yo24} and \citet{yo25a} to model sub-millimetre flux densities. Specifically, we assume that $40\%$ of the metals in the cold gas phase of the galaxies are in the form of dust \citep{dwek98}, which is a reasonable assumption for massive galaxies. The sub-millimetre flux density at $870\,\mu\rm{m}$ ($S_{870}$) is then estimated using the scaling relation presented by \citet{rachel23}, which is derived using high-resolution radiative transfer simulations \citep[see][]{cochrane19,cochrane22,cochrane23,cochrane_dark}.
Essentially, in their relation, the $S_{870}$ flux density depends on the star-formation rate (SFR), stellar mass ($M_{\star}$), dust mass ($M_{\rm dust}$), and redshift ($z$) as follows: 
\begin{equation} \label{cochrane_relation}
%\small
    \frac{S_{870}}{\rm mJy} = \alpha \left (  \frac{\rm SFR}{100 M_{\odot} \ {\rm yr}^{-1}}\right )^{\beta} \left (  \frac{M_{\star}}{10^{10} M_{\odot}}\right )^{\gamma} \left (  \frac{M_{\rm dust}}{10^{8} M_{\odot}}\right )^{\delta} (1 + z)^{\eta},
\end{equation}
where $\log \alpha = -0.77$, $\beta = 0.32$, $\gamma=0.13$, $\delta=0.65$, and $\eta=0.65$. We note that the SFR is calculated as the total stellar mass formed within the last $\sim$14 Myr (a timescale that decreases with increasing redshift), consistent with the interval adopted in \citet{cochrane23}. This timescale corresponds to the internal binning of 20 sub-steps that \texttt{L-Galaxies} uses to trace galaxy evolution between two consecutive snapshots. 

\begin{figure*}
    \centering
    \includegraphics[width=\textwidth]{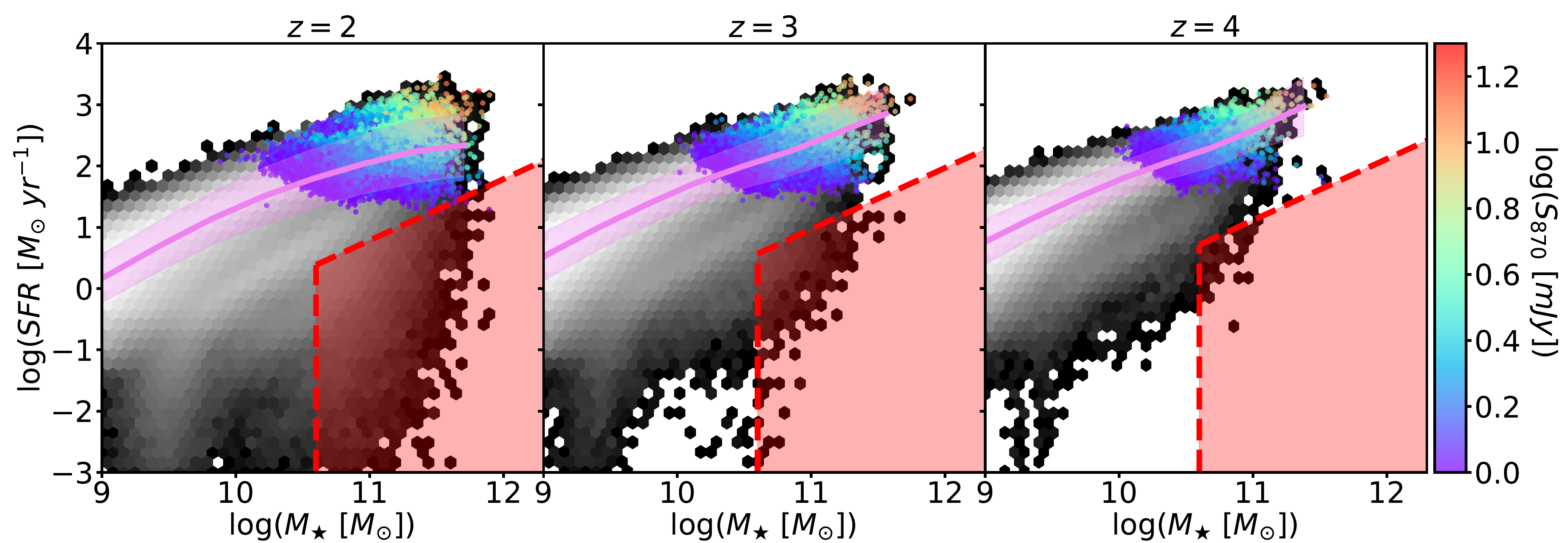}
    \caption{The star formation rate (SFR)–stellar mass ($M_{\star}$) relation at $z = 2$, 3, and 4 is shown in the first, second, and third panels, respectively, using the re-calibrated version of \texttt{L-Galaxies} from \citet{yo25a}. The coloured dots represent bright DSFGs with $S_{870} \geq 1 \,\rm{mJy}$, colour-coded by their $S_{870}$ flux density (see colour-bar), derived using the scaling relations presented by \citet{rachel23}. The red dashed lines indicate the thresholds in specific star formation rate (sSFR) and stellar mass used to separate quenched and massive galaxies, respectively, following the definition of \citet{carnall20}. The pink line and coloured area correspond to the `main sequence' of galaxies in our model and its $1\sigma$ dispersion.
    Overall, bright DSFGs are located in the massive, star-forming region of the diagram and lie on, or just above, the `main sequence'. As expected, the massive, quenched region is less populated at higher redshifts.}
    \label{fig:SFR-Mstar}
\end{figure*}

\section{Results} \label{sec:results}
In this section, we present the evolution of key galaxy properties across cosmic time for the massive quiescent galaxies within our semi-analytic model. In particular, we aim to explore the link between this population and sub-millimetre-bright dusty star-forming galaxies.

In Figure \ref{fig:SFR-Mstar}, we present the star-formation rate (SFR)–stellar mass ($M_{\star}$) relation of galaxies in our model at similar redshifts as to the selected MQs analysed in this work. We define MQs following the criteria of \citet{carnall20}: 
\begin{itemize} 
\item $\log (M_{\star}/\rm{M_{\odot}}) \geq 10.6$ 
\item ${\rm{sSFR}} \leq 0.2 / t_{{\rm obs}} (z)$ 
\end{itemize} 
where sSFR denotes the specific star-formation rate (SFR$/M_{\star}$), and $t_{\rm obs} (z)$ is the age of the Universe at redshift $z$. The MQ regime is indicated in Figure \ref{fig:SFR-Mstar} by the light red shaded region. As expected, the number of MQs in our model declines with increasing redshift. In total, we find $10,805$ MQs at $z = 2$, $446$ at $z = 3$, and just $21$ at $z = 4$ within the full $(713.6 \, {\rm cMpc})^3$ \texttt{Millennium} volume. For completeness, we also show the `main sequence' in Figure \ref{fig:SFR-Mstar}, which we obtained by calculating, in stellar mass bins, the median and standard deviation of the SFR of star-forming (${\rm sSFR} > 0.2 / t_{{\rm obs}} (z)$) galaxies.

We model the $870\,\mu\rm{m}$ sub-millimetre flux density ($S_{870}$) based on each galaxy's dust mass, star formation rate, stellar mass and redshift, using the scaling relations presented by \citet{rachel23}, as described above. It is important to note that the default \citet{henriques20} version of \texttt{L-Galaxies} overpredicts cold-gas metallicities compared to observational data at high redshift (by $\sim 0.40 \, {\rm dex}$ for galaxies with $9.0 \leq \log(M_{\star} / M_{\odot}) < 10$ at $z \sim 3$). This issue was addressed by \citet{yates21}, who increased the metal-ejection efficiency from SN feedback (free parameters not included in the MCMC framework of \citet{henriques20} or \citet{yo25a}). We find that our recalibrated version predicts cold-gas metallicities that are slightly higher (by $\sim 0.25 \, {\rm dex}$ for galaxies in the same stellar-mass and redshift bins) than those of \citet{yates21}.
We classify DSFGs as those with $S_{870} \geq 1\,\rm{mJy}$, in line with the faintest detected sources in large sub-millimetre surveys such as AS2UDS \citep{simpson17, stach18, stach19, dud20}. We show their positions in the SFR–$M_{\star}$ plane as coloured dots, where the colour encodes the sub-millimetre flux. From Figure \ref{fig:SFR-Mstar}, we see that most DSFGs in our model are already massive at the selected redshifts. Brighter DSFGs tend to populate the higher-mass end of the plane, residing on or above the main sequence. Notably, all DSFGs with $S_{870}\gtrsim2\,\rm{mJy}$ exceed the stellar mass threshold used to define MQs in this study. The formation and evolution of DSFGs at $z = 2$, $3$, and $4$ will be studied more extensively in a future work. Here, we will focus primarily on the progenitors of MQs.

\begin{figure*}
    \centering
    \includegraphics[width=\textwidth]{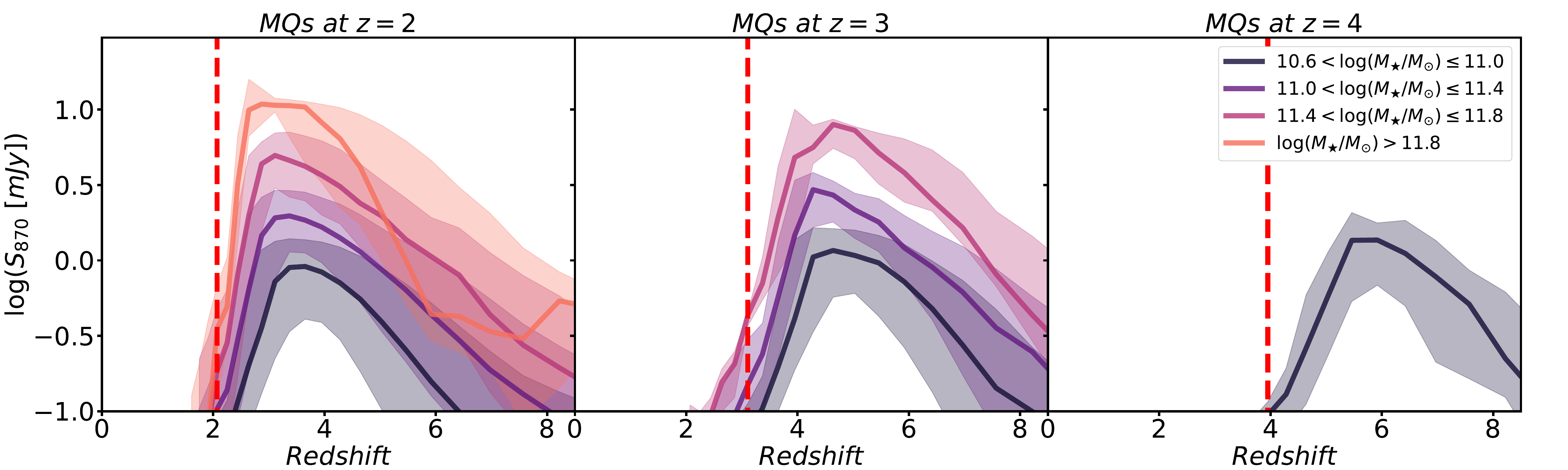}
    \caption{The evolution of the $S_{870}$ flux density of massive quiescent galaxies (MQs) selected at $z = 2$, 3, and 4 is shown in the first, second, and third panels, respectively. At each selected redshift, MQs are divided into four stellar mass bins of width $0.4\,\rm{dex}$, starting from a lower limit of $\log(M_{\star}/\rm{M_{\odot}}) = 10.6$. The coloured regions and lines represent the $16^{\rm{th}}$, $50^{\rm{th}}$, and $84^{\rm{th}}$ percentiles of MQs within each stellar mass bin. Overall, the $S_{870}$ evolution of MQs exhibits a gradual increase with decreasing redshift, reaching a peak before undergoing a rapid decline at later times. Notably, the most massive MQs consistently display the highest $S_{870}$ peak values.}
    \label{fig:S870-track_MQs}
\end{figure*}

\begin{figure*}
    \centering
    \includegraphics[width=\textwidth]{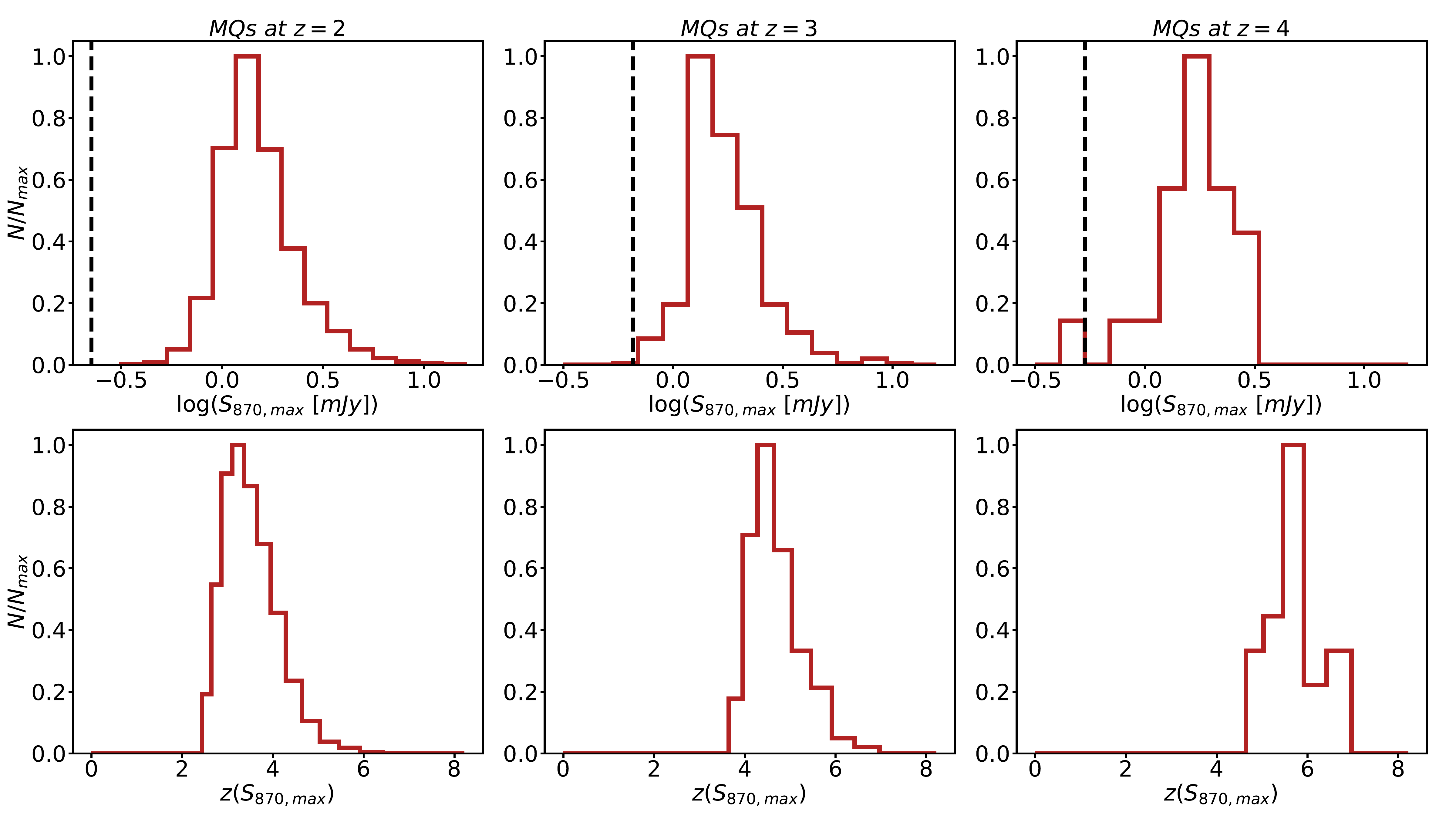}
    \caption{The distribution of the maximum $S_{870}$ flux density reached by each modelled galaxy across its entire formation history, $S_{870,\mathrm{max}}$ (first row), and when it occurred (second row; redshift distribution), for massive quiescent galaxies (MQs) selected at $z = 2$, $3$, and $4$ (from left to right, respectively). The minimum $S_{870,\mathrm{max}}$ reached by any of the MQs is indicated by the vertical dashed black line.}    
    \label{fig:S870-dist}
\end{figure*}

\subsection{High-redshift massive quiescent galaxies were sub-millimetre bright at earlier cosmic times} \label{sec:results_1}
In this section, we explore possible evolutionary connections between MQs and DSFGs. We seek to answer the following questions: were MQs sub-millimetre-bright before quenching?; what was the maximum sub-millimetre flux density reached by MQs in the past, and when did this peak occur?; do the answers to these questions depend on the redshift at which the MQ is identified?\\
\indent As described above, we select massive MQs at $z = 2$, $3$, and $4$. In order to analyse the evolution of these selected galaxies across cosmic time, we track each galaxy along its so-called Main Progenitor Branch (MPB) by matching galaxies at different redshifts with the same \texttt{MainLeafID} (i.e. that have the same original progenitor galaxy). First, we explore the past evolution of sub-millimetre flux density for our MQ samples selected at different redshifts. In Figure \ref{fig:S870-track_MQs}, we show the median evolution of $S_{870}$ flux density for MQs selected at $z = 2$, $3$, and $4$ in four stellar mass (at the selected redshift) bins. The most notable trend in Figure \ref{fig:S870-track_MQs} is that MQs tend to reach a peak in sub-millimetre flux density before the redshift at which they are selected. The typical $S_{870}$ evolution exhibits a gradual increase with cosmic time (i.e. with decreasing redshift), between $z \sim 9$ and its peak, followed by a rapid decline at lower redshifts. This can be seen for MQs at the three analysed redshifts. Another noteworthy trend is that, on average, MQs with higher stellar masses tend to reach higher peak sub-millimetre flux densities. \\
\indent Motivated by the key trend in Figure \ref{fig:S870-track_MQs}, where MQs selected at different redshifts exhibit a peak in sub-millimetre flux density at different cosmic times, we derive both the maximum sub-millimetre flux density, $S_{870, \mathrm{max}}$, and the redshift at which it occurs, $z(S_{870, \mathrm{max}})$, for every MQ. The distributions of these properties are shown in Figure \ref{fig:S870-dist}, for each of the three MQ samples. As anticipated from Figure \ref{fig:S870-track_MQs}, almost all MQs at the three selected redshifts exhibit bright sub-millimetre peaks ($S_{870, \mathrm{max}} \gtrsim 1 \, {\rm mJy}$). We see from the upper panel of Figure \ref{fig:S870-dist} that $\sim 86\%$, $96\%$, and $91\%$ of MQs at $z = 2$, $3$, and $4$, respectively, exceeded $1\,{\rm mJy}$ at some point in their evolution. However, only $\sim 7\%$, $9\%$, and $5\%$ of these MQs (at the same redshifts) reached $S_{870,{\rm max}} \geq 3\,{\rm mJy}$. In exceptional cases, $\sim 2\%$, $1\%$, and $0\%$ MQs at $z = 2$, $3$, and $4$, attained $S_{870} \geq 5 \,{\rm mJy}$ during their formation history. Therefore, in our model, the progenitors of the vast majority of MQs are DSFGs. The median $S_{870, \mathrm{max}}$ is $1.4$, $1.6$, and $1.8\,\rm{mJy}$, for MQs at $z = 2$, $3$, and $4$. Based on the redshift distributions displayed in the lower panel of Figure \ref{fig:S870-dist}, these sub-millimetre peaks occurred at the following median redshifts: $z(S_{870, \mathrm{max}}) = 3.1$, $4.3$, and $5.5$, for MQs selected at $z =2$, $3$, and $4$, respectively. \\
\begin{figure*}
    \centering
    \includegraphics[width=\textwidth]{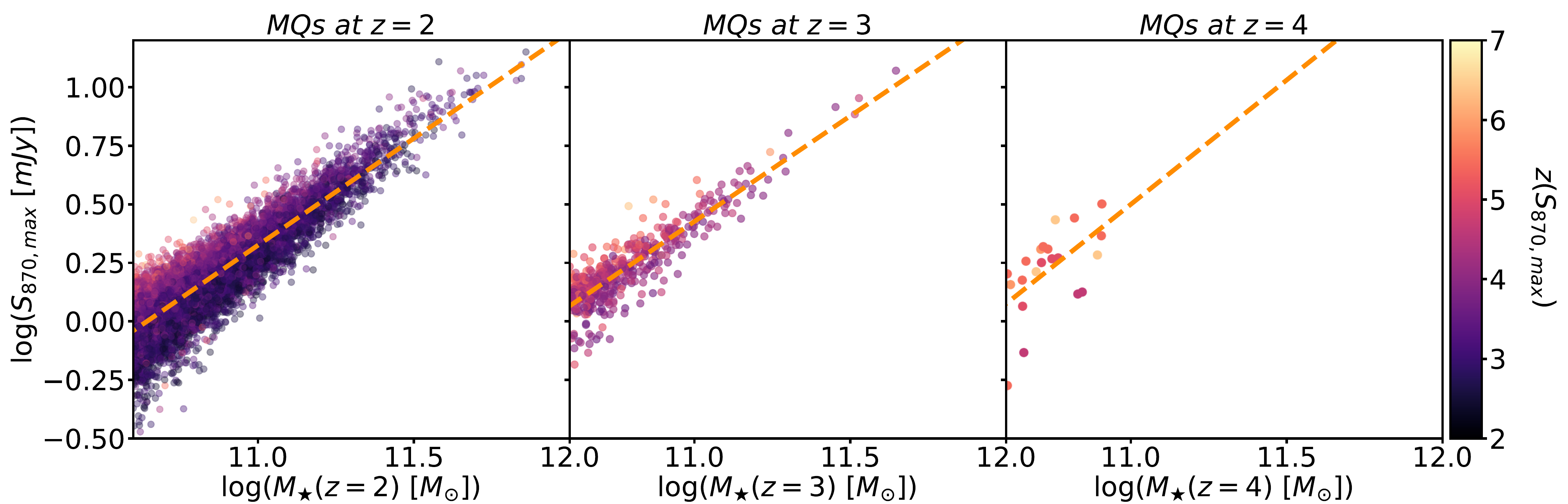}
    \caption{The maximum value of $S_{870}$, $S_{870,\mathrm{max}}$, reached across the formation history of massive quiescent galaxies (MQs), as a function of stellar mass at redshifts $z = 2$, $3$, and $4$ (from left to right). The data points are coloured according to the redshift at which $S_{870,\mathrm{max}}$ occurred, $z(S_{870,\mathrm{max}})$. A strong correlation is found between the stellar mass of MQs and their $S_{870,\mathrm{max}}$, with a dispersion that also correlates with $z(S_{870,\mathrm{max}})$. At fixed stellar mass, MQs that reached $S_{870,\mathrm{max}}$ at earlier times tended to exhibit brighter $S_{870}$ flux densities.}
    \label{fig:S870-smass}
\end{figure*}
\indent Another tentative correlation apparent in Figure~\ref{fig:S870-track_MQs} is that more massive MQs tend to have reached higher peak sub-millimetre fluxes. To explore this further, we present in Figure~\ref{fig:S870-smass} the maximum $870\,\mu\rm{m}$ flux density reached by MQs ($S_{870, {\rm max}}$) as a function of their stellar mass at the selected redshift. Our results reveal a strong correlation between these two properties across all three redshifts analysed. The scatter in this relation is typically less than $\sim 0.5\,\mathrm{dex}$ and tends to decrease with increasing stellar mass. Interestingly, at fixed stellar mass, the deviation from the median relation is correlated with the redshift at which the sub-millimetre peak occurred: MQs with a given stellar mass, $M_{\star}$, reached a brighter $S_{870, {\rm max}}$ if the sub-millimetre emission peaked at an earlier epoch. This correlation arises because, in our model, galaxies that reached their sub-millimetre peak earlier have, on average, higher star-formation rates than those at lower redshifts.\\
\indent The correlation between stellar mass and $S_{870, {\rm max}}$ appears to be independent of the redshift at which the MQs are selected. To quantify this, we perform a linear fit between $S_{870, {\rm max}}$ and stellar mass for each sample. 
In Table~\ref{tab:S870-smass}, we present the best-fit parameters and their associated uncertainties in for the linear function:
\begin{equation} 
\log (S_{870,{\rm max}}) = a  \log( M_{\star}/\rm{M_{\odot}}) + b
\label{eq:linear-fit}
\end{equation}

\begin{table}
\centering
\caption{The best-fit parameters for Equation \ref{eq:linear-fit}} 
\label{tab:S870-smass}
\begin{tabular}{ccc}
\hline
$z ({\rm MQ})$ & $a$ & $b$ \\
\hline
$2$ & $0.913\pm 0.004$ & $-9.714\pm 0.047$ \\
$3$ & $0.901\pm 0.021$ & $-9.488\pm 0.231$\\
$4$ & $1.061 \pm 0.350$ & $-11.173\pm 3.761$ \\
\hline
\end{tabular}
\tablefoot{Equation \ref{eq:linear-fit} relates the maximum sub-millimetre flux-density reached by massive quiescent galaxies selected at $z({\rm MQ})$, across their history ($S_{870, {\rm max}}$) to the stellar mass at $z({\rm MQ})$.}
\end{table}
\noindent Based on the best-fit parameters, the derived scaling relation is, within the uncertainties, redshift-independent. This relation could thus be used to estimate the maximum $870\,\mu\rm{m}$ flux density reached by massive quiescent galaxies at high redshift, at least within the redshift range $z=2-4$.\\
\begin{figure*}
    \centering
    \includegraphics[width=\textwidth]{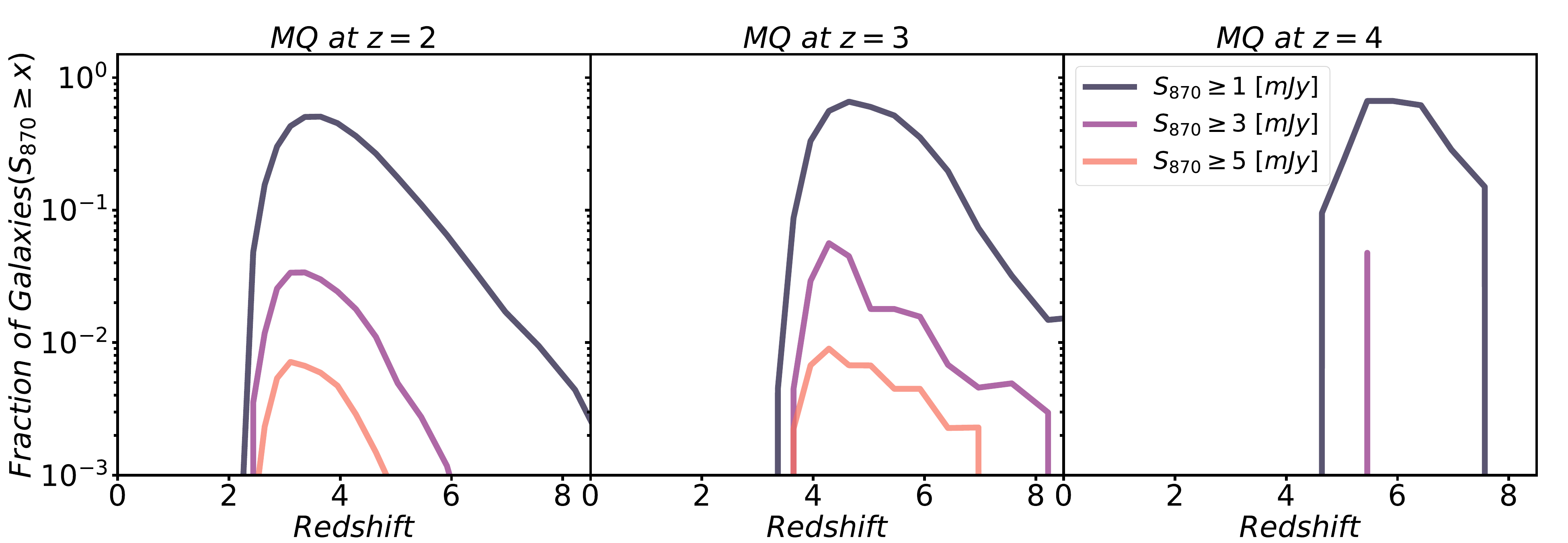}
    \caption{The fraction of massive quiescent galaxy (selected at $z = 2$, $3$, and $4$; first, second, and third panels, respectively) progenitors that exceed $S_{870}$ flux densities of $1\,\rm{mJy}$ (dark blue), $3\,\rm{mJy}$ (purple), and $5\,\rm{mJy}$ (orange), versus redshift. At the redshift at which the distributions peak, a significant fraction of MQ progenitors exceed $S_{870}=1\,\rm{mJy}$. For example, $50\%$, $66\%$, and $67\%$ of MQs selected at $z=2$, $3$, and $4$, respectively, exceeded $S_{870}=1\,\mathrm{mJy}$ at $z \sim 3.4$, $4.3$, and $5.5$. These numbers are lower for higher $S_{870}$ thresholds: at their peaks, $\sim 3\%$ of MQs selected at $z=2$, $\sim 6\%$ at $z=3$, and $\sim 5\%$ at $z=4$ exceed $3\,\rm{mJy}$.}
    \label{fig:Fraction-S870-z}
\end{figure*}
\indent The bottom panels of Figure~\ref{fig:S870-dist} show the distribution of redshifts at which individual MQs reached their sub-millimetre peak, $z(S_{870, {\rm max}})$. However, the median $z(S_{870, {\rm max}})$ is not necessarily the redshift at which the majority of MQs were sub-millimetre bright. For example, some MQs already surpass $S_{870} \geq 1 \, {\rm mJy}$ prior to their $z(S_{870, {\rm max}})$, as is shown in Figure \ref{fig:S870-track_MQs} (see the gradual increase of $S_{870}$ with decreasing redshift, prior to the peak). To assess this, in Figure~\ref{fig:Fraction-S870-z} we quantify the fraction of MQs selected at $z = 2$, $3$, and $4$ that exceeded a given $S_{870}$ threshold ($1\,\rm{mJy}$, $3\,\rm{mJy}$, or $5\,\rm{mJy}$; see three coloured lines), as a function of redshift. The redshift at which the maximum number of the main progenitors of MQs exceeded $S_{870} = 1\,\rm{mJy}$ is $z \sim 3.4$, $4.3$, and $5.5$, for MQs selected at $z = 2$, $3$, and $4$, respectively. Overall, the redshifts at which most MQs exceeded the three flux thresholds are broadly consistent with the redshifts at which they reached $S_{870, {\rm max}}$. At these peaks, a significant fraction — $50\%$, $66\%$, and $67\%$ — of the main progenitors of MQs selected at $z = 2$, $3$, and $4$, exceed the threshold of $S_{870} = 1\,\rm{mJy}$. However, only a small fraction (less than $\sim 8\%$) ever reached flux densities brighter than $3\,\rm{mJy}$. For example, just one  ($\sim$ 5 \%) MQ selected at $z = 4$ exceeded $S_{870} = 3 \, \mathrm{mJy}$, and none surpassed $5 \,\mathrm{mJy}$. 

\subsection{Not all dusty star-forming galaxies are progenitors of high-redshift MQs}

\begin{figure*}
    \centering
    \includegraphics[width=\textwidth]{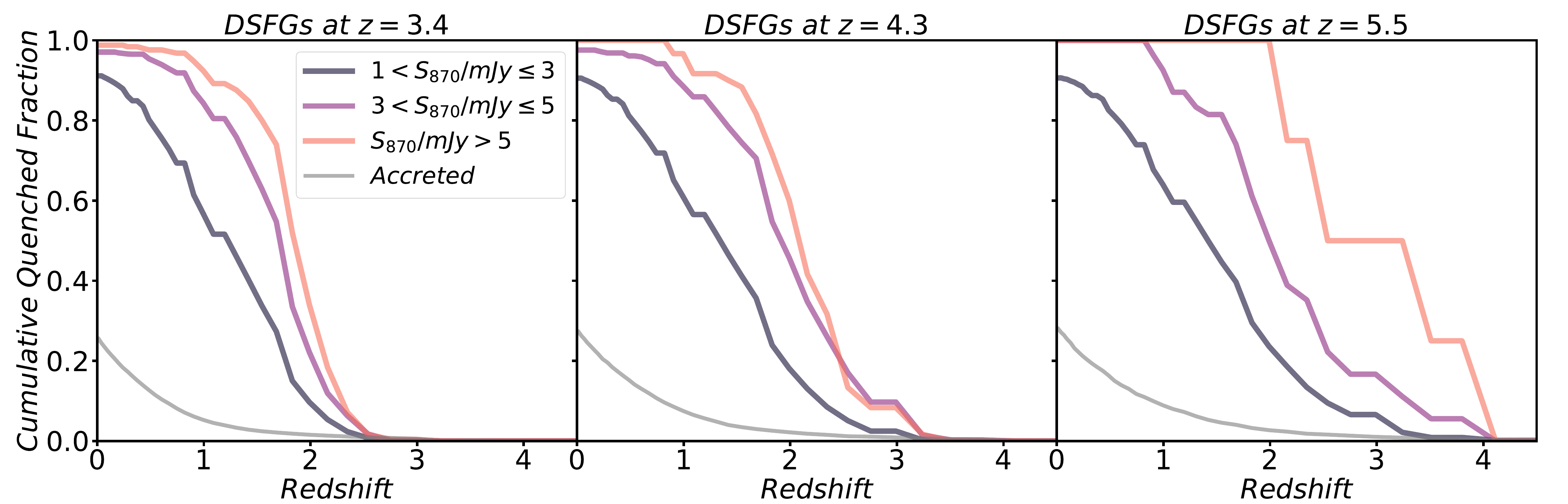}
    \caption{The cumulative fraction of dusty star-forming galaxies (DSFGs) selected at $z = 3.4$, $4.3$, and $5.5$ (from left to right) that become massive and quiescent, as a function of redshift. The dark blue, purple, and orange lines represent DSFGs with sub-millimetre flux densities of $1 < S_{870}/\rm{mJy} \leq 3$, $3 < S_{870}/\rm{mJy} \leq 5$, and $S_{870}/\rm{mJy} > 5$, respectively. The grey line indicates fraction of DSFGs that were accreted by a more massive galaxy. We note that all DSFGs with $S_{870} \gtrsim 3 \, \rm{mJy}$ are already massive at all selection redshifts. The most remarkable trend is that brighter DSFGs quench more rapidly than their fainter counterparts. For example, $50\%$ of the bright DSFGs selected at $z = 3.4$, $4.3$, and $5.5$ become quenched by $z \sim 1.85$, $2.08$, and $2.54$, respectively. In contrast, the fainter DSFGs across all selected redshifts tend to quench later, with similar `quenched redshift', typically around $z \sim 1.3$. This is because faint DSFGs selected at high redshift may continue to grow and potentially become brighter at later times.}
    \label{fig:Fraction-SMG-Q}
\end{figure*}

We have already shown that, in our model, massive quiescent galaxies at $z = 2$, 3, and 4 were sub-millimetre bright in the past. For instance, we find that $\sim 86\%$, $96\%$, and $91\%$ of MQs at $z = 2$, 3, and 4, exceeded $1\,{\rm mJy}$ at some point in the past. Only $\sim 7\%$, $9\%$, and $5\%$ reached $S_{870,{\rm max}} \geq 3\,{\rm mJy}$, while $\sim 2\%$, $1\%$, and $0\%$ surpassed $5\,{\rm mJy}$. However, it is not yet clear whether a significant fraction of all DSFGs -- defined here as galaxies with $S_{870} \geq 1\,{\rm mJy}$ -- at those redshifts are indeed progenitors of MQs at $2 \leq z \leq 4$. For simplicity, we do not follow galaxies once they are accreted by a more massive system. However, the fraction of MQs that are eventually incorporated into a more massive galaxy is modest ($\lesssim24\%$), regardless of the selected redshift. \\
\indent In this section, we analyse the entire DSFG population at the redshifts when the sub-millimetre emission of the selected MQs at $z = 2$, $3$, and $4$ was most common. We therefore select DSFGs at $z = 3.4$, $4.3$, and $5.5$, respectively (based on Figure \ref{fig:Fraction-S870-z}). We separate the DSFG sample into three bins based on sub-millimetre flux density at the selected redshift: faint ($1 < S_{870}/{\rm mJy} \leq 3$), intermediate ($3 < S_{870}/{\rm mJy} \leq 5$), and bright ($S_{870}/{\rm mJy} > 5$). At $z = 3.4$, these bins include $56,753$, $1,500$, and $250$ DSFGs. The corresponding numbers are $22,771$, $411$, and $60$ at $z = 4.3$, and $5,460$, $54$, and $4$ at $z = 5.5$. Already, we note that the number of DSFGs is significantly larger than that of MQs at $z = 2$, $3$, and $4$. Thus, we expect that not all DSFGs in our model are progenitors of MQs at $2 \lesssim z \lesssim 4$. Next, we will quantify how many of the overall DSFG population evolve into high-redshift MQs and when these DSFGs quench.\\
\indent In Figure~\ref{fig:Fraction-SMG-Q}, we show the cumulative fraction of DSFGs — binned by $S_{870}$ — that become massive ($\log(M_{\star}/\rm{M_{\odot}}) \geq 10.6$) and quiescent (sSFR $\leq 0.2/t_{\rm obs}(z)$). Notice that cumulative fractions for faint DSFGs never reach 100\%, primarily because some of them are accreted into more massive systems. Figure~\ref{fig:Fraction-SMG-Q} reveals that brighter DSFGs quench earlier. For example, all bright DSFGs ($S_{870}/{\rm mJy} > 5$) identified at $z = 5.5$ become massive and quiescent by $z = 2$, whereas only $\sim 40\%$ of intermediate and $\sim 20\%$ of faint DSFGs do. We also calculate the redshift at which $50\%$ of bright DSFGs selected at $z = 3.4$, $4.3$, and $5.5$ become massive and quiescent. These are $z_{\rm MQ} = 1.85$, $2.08$, and $2.54$, respectively. In contrast, 50\% of faint DSFGs ($1 < S_{870}/{\rm mJy} \leq 3$) selected at the same redshifts meet the MQ criteria only by $z_{\rm MQ} = 1.23$, $1.34$, and $1.42$, respectively. In summary, DSFGs that are higher-redshift and brighter quench earlier.\\
\indent We reiterate that the majority of MQs at $z = 2$, $3$, and $4$ lie within the faintest  $S_{870}$ bin (see Figures \ref{fig:S870-dist} and \ref{fig:Fraction-S870-z}). However, less than 50\% of faint DSFGs at $z = 5.5$ will become MQs by $z = 2$. Therefore, while the majority of MQs at $z = 2$, $3$, and $4$ were DSFGs at $z = 3.4$, $4.3$, and $5.5$, only a small fraction of all DSFGs evolve into massive and quiescent by those redshifts ($z \gtrsim 2$). This implies that MQs may follow a distinct evolutionary path, different from that of the average DSFG.\\
\begin{figure*}
    \centering
    \includegraphics[width=\textwidth]{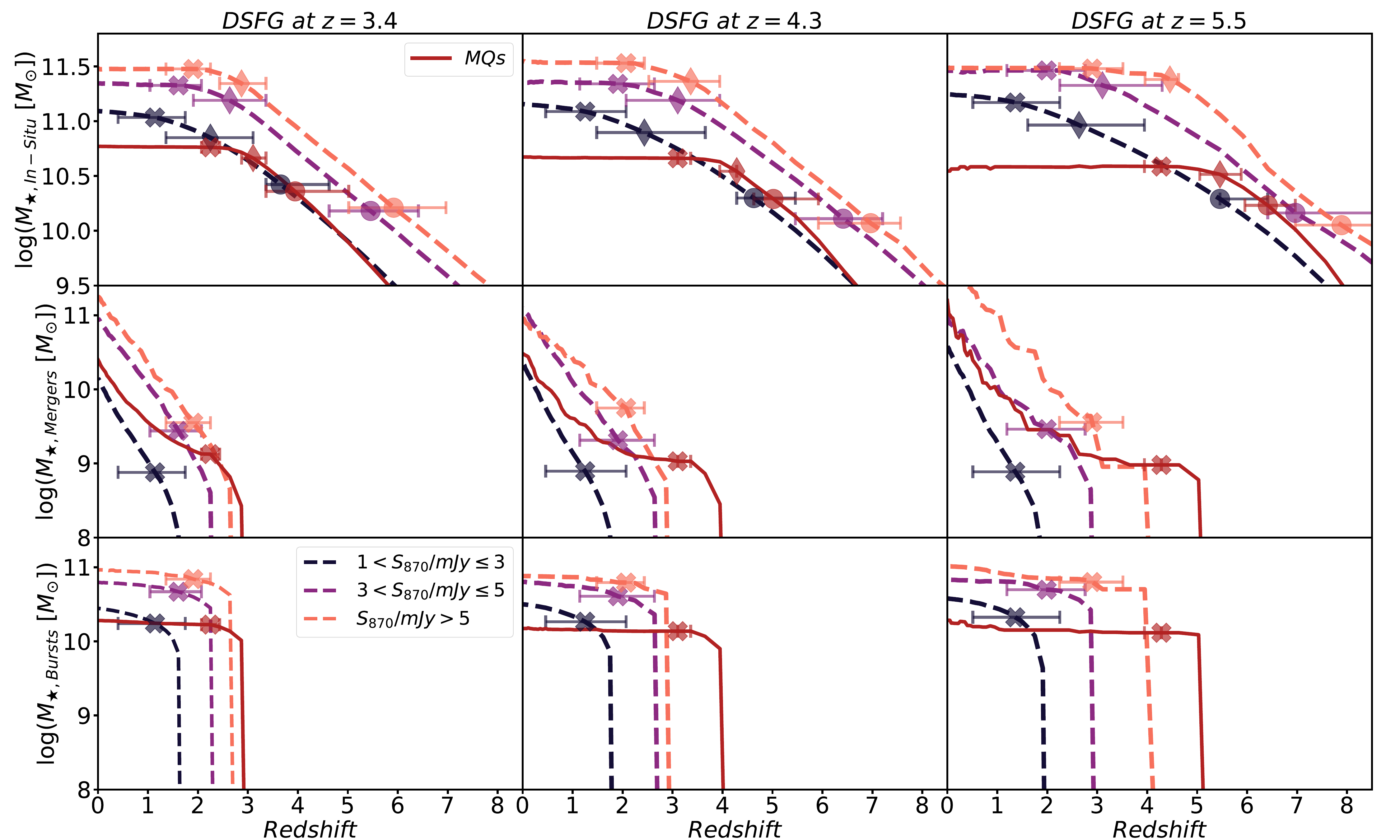}
    \caption{The median stellar mass evolution attributed to secular processes (In-Situ; top panels), accretion via mergers (Mergers; middle panels), and merger-induced starbursts (Burst; bottom panels) for massive quiescent progenitors (solid red line) the bright (dashed orange line), intermediate (purple dashed lines), and faint DSFGs (dark blue dashed lines). From left to the right, we show DSFG and MQ progenitors selected at $z = 3.4$, $4.3$, and $5.5$, respectively. The dots, diamonds, and crosses  highlight the median redshifts at which the galaxy population exceeded $S_{870} > 1 \, {\rm mJy}$ for the first time, reaches the sub-millimetre peak, and quenches, respectively. The formation of all four galaxy populations is initially dominated by secular star formation, suggesting that sub-millimetre emission is not primarily triggered by mergers at early times. The main difference in stellar mass growth between typical DSFGs and MQ progenitors lies in the timing of major mergers. MQ progenitors experience a significant merger event earlier, which induces an extreme starburst and rapid SMBH-driven cold gas accretion, thereby depleting the available gas for continued secular star formation. Following this early, violent merger, these galaxies grow their stellar mass primarily through dry mergers.}
    \label{fig:m_channels}
\end{figure*}
\indent To highlight the main differences in the evolution of high-redshift MQ progenitors compared to the overall DSFG population, we present in Figure \ref{fig:m_channels} the median stellar mass growth attributed to secular processes (In-Situ), accretion via major and minor mergers (Mergers), and merger-induced starbursts (Burst) for bright, intermediate, and faint DSFGs, compared with the MQ population. To aid interpretation, Figure \ref{fig:m_channels} also shows the median redshifts at which the galaxy population exceeded $S_{870} > 1 \, {\rm mJy}$ for the first time (dots), reach their sub-millimetre flux peak (diamonds), and quench (crosses). Figure \ref{fig:m_channels} shows that the dominant mechanism responsible for building up stellar mass in DSFGs at early epochs ($z \gtrsim3$) is secular star formation. In general, the slope of the in-situ stellar mass growth is similar across all galaxy populations, with differences primarily in their initial stellar mass. A notable exception is the brightest DSFGs and MQ progenitors selected at $z = 5.5$, which exhibit accelerated in-situ growth at $z \sim 6$ for bright DSFGs and $z \sim 7$ for MQ progenitors. By contrast, MQ progenitors selected at lower redshifts ($z = 3.4$ and $4.3$) follow similar in-situ mass growth to faint DSFGs selected at the same redshifts. This is consistent with the finding that most MQs selected at $z = 2$, $3$, and $4$ reached an $S_{870, {\rm max}}$ of between $1\,\rm{mJy}$ and $3\,\rm{mJy}$ (see Figure \ref{fig:S870-dist}).\\
\indent The redshift at which in-situ stellar mass growth begins to flatten — indicating suppression of star formation — coincides with the redshift at which stellar mass accretion from mergers (and associated starbursts) becomes significant across all four galaxy populations. On average, this transition occurs at a slightly lower redshift than the selection redshift of the MQs, and after the typical redshift when they reached their sub-millimetre emission peak. This suggests that, on average, these galaxies became sub-millimetre bright primarily due to secular processes rather than merger-induced starbursts. It is important to note that the redshift at which $S_{870, {\rm max}}$ occurs is not precisely determined, due to the time resolution of the simulation ($\sim 14$ Myr). Nevertheless, these galaxies are already sub-millimetre bright prior to reaching their maximum sub-millimetre flux density.\\
\indent The primary difference in stellar mass evolution across the four populations is when the first significant merger event occurred. Figure \ref{fig:m_channels} clearly shows that violent mergers occurred earlier in MQ progenitors than in the broader DSFG population, setting them on distinct evolutionary paths. In fact, on average, the progenitors of MQs selected at $z=2$ and 3, and faint DSFGs (at $z = 3.3$ and 4.3), exceeded the $S_{870} = 1\,{\rm mJy}$ threshold at similar epochs, whereas the progenitors of MQs at $z=4$ align better with intermediate DSFGs at $z=5.5$. In all cases, the sub-millimetre episode occurred before the violent merger. Therefore, galaxies become DSFGs ($S_{870} > 1 \; {\rm mJy}$) via in-situ star formation.
Among the broader DSFG population, brighter DSFGs experienced mergers earlier than fainter ones. For example, for DSFGs selected at $z = 5.5$ (when sub-millimetre emission was prevalent among MQs at $z = 4$), the median redshift of the first major merger event is approximately $z \sim 5$ for those DSFGs that are MQ progenitors, and $z=4$, $3$, and $2$ for the wider bright, intermediate, and faint population of DSFGs, respectively.\\
\indent Following the first major merger, stellar mass growth is dominated by additional mergers, often without significant subsequent starbursts, suggesting dry merger events. Notably, while MQ progenitors underwent at least one significant merger earlier, their stellar mass accretion via mergers shows an extended plateau phase (seen particularly clearly for those selected at $z = 5.5$ and $4.3$), indicating a period with few significant mergers. This is followed by a renewed phase of growth from mergers, though at a slower rate, on average, than is seen in DSFGs (except for MQ progenitors at $z = 5.5$, whose post-plateau mass growth is similar to that of the wider population of intermediate-luminosity DSFGs).\\
\begin{figure*}
    \centering
    \includegraphics[width=\textwidth]{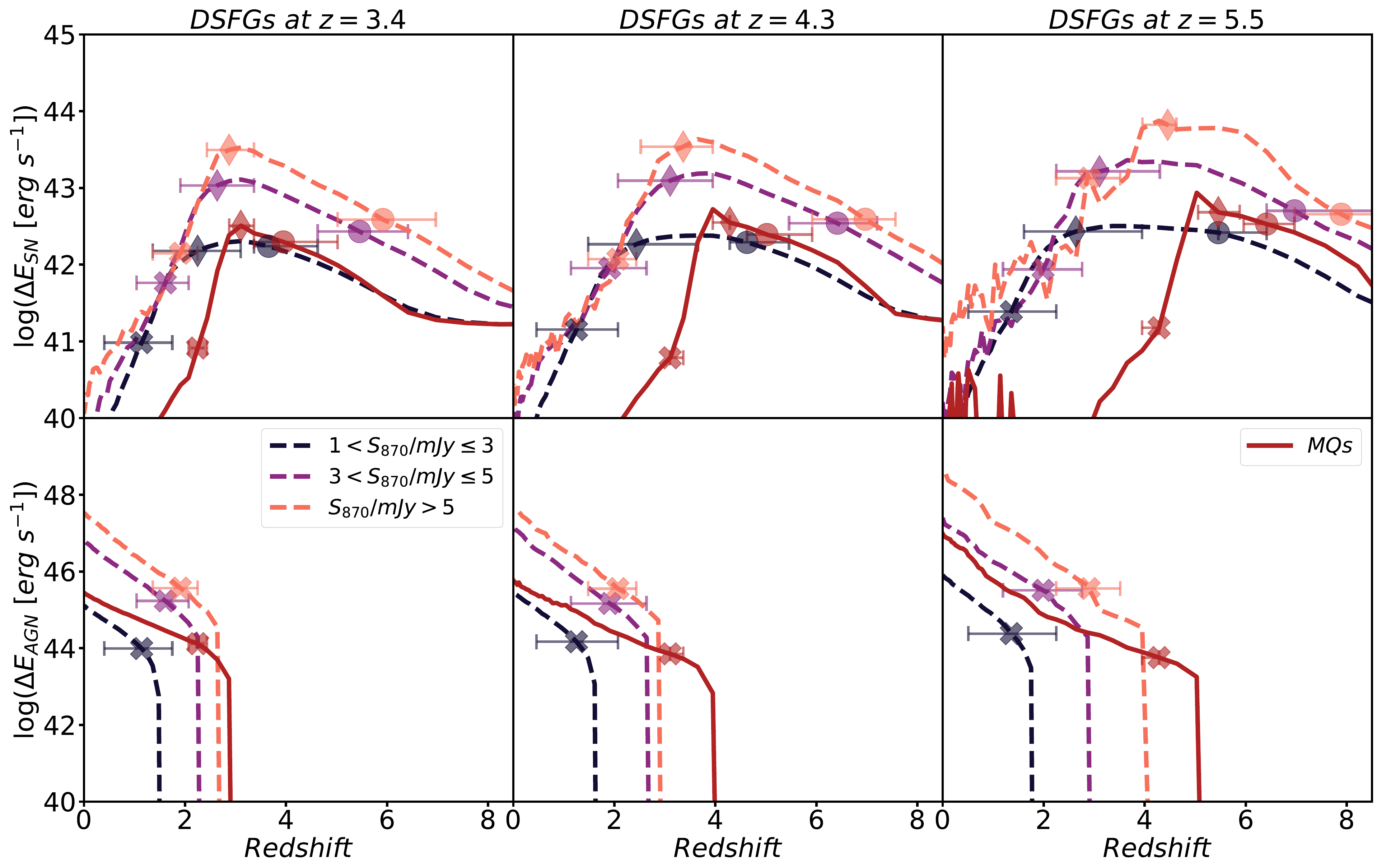}
    \caption{The evolution of the median energy input rate from supernovae (SN; top panels) and active galactic nuclei (AGN; bottom panels) for massive quiescent progenitors (red solid lines), faint (dark blue dashed lines), intermediate (purple dashed lines), and bright (orange dashed lines) dusty star-forming galaxies selected at $z = 3.4$ (left-hand panel), $z=4.3$ (central panel), and $z=5.5$ (right-hand panel) -- the redshifts at which the majority of MQs at $z=2$, $3$, and $4$, were DSFGs. The crosses indicate the median redshift at which a given galaxy population first satisfied the quiescent criteria (sSFR $\leq 0.2/t_{\rm obs}(z)$). The energy input from AGN feedback rises rapidly close to the quenching epoch, dominating the energy injection rate over SN feedback by at least $2$ orders of magnitude. This suggests that AGN feedback is the main mechanism that quenches the star formation in the four populations, but that this happens earlier for MQs.}
    \label{fig:feedback}
\end{figure*}
\indent Figure \ref{fig:m_channels} also indicates that the first significant merger triggered an intense starburst (see lower panels) that consumed a significant amount of cold gas. However, this was not the unique mechanism responsible for quenching the star formation, since a brief period of secular star formation followed. As previously discussed, in the \texttt{L-Galaxies} SAM, mergers are the main channel for supermassive black hole (SMBH) growth via cold gas accretion. In particular, the model employed in this work assumes highly efficient accretion, leading to rapid SMBH growth. Consequently, merger events can simultaneously drive both extreme starbursts (consuming cold gas) and AGN activity (suppressing cooling).\\
\indent To investigate the primary physical processes responsible for quenching star formation in DSFG and MQ progenitors, we explore the evolution of the median energy injection rates due to supernova (SN) and AGN feedback\footnote{ see Appendix \ref{sec:energy-inputs} for a detailed description of how these quantities are calculated.} for the four galaxy populations at the three selected redshifts, as shown in Figure \ref{fig:feedback}. The most prominent trend in Figure \ref{fig:feedback} is that the SN feedback energy injection rate gradually increases from early times ($z \sim 9$), peaking at a redshift that coincides with the onset of AGN feedback — which also matches the average redshift of the first relevant merger (see Figure \ref{fig:m_channels}). After this peak, the SN energy input rapidly declines, a trend that can be attributed to the drop in star formation following the intense merger-induced starburst, particularly in MQ progenitors.\\
\indent The lower panels of Figure \ref{fig:feedback} show that quenching (see cross symbols) typically occurs after strong AGN feedback. While SN feedback continues to contribute energy, its input is, on average, more than two orders of magnitude lower than that from AGN (see upper panels). Hence, the AGN feedback appears to be the dominant mechanism responsible for quenching in DSFGs and MQs in our model — particularly when triggered early. In the \texttt{L-Galaxies} framework, AGN feedback operates by injecting energy into the hot gas halo, thereby suppressing cooling. The AGN energy injection rate is tied to hot gas accretion onto the SMBH, which scales with both the hot gas mass and the SMBH mass. This means that significant AGN activity is initiated in galaxies with large hot gas reservoirs and/or large SMBHs grown through mergers and cold gas accretion. This is coupled with the starburst, which consumes a large fraction of the cold gas and reduces the star formation rate. The subsequent decline in SN feedback energy is therefore a natural consequence of depleting star formation. While Figure \ref{fig:feedback} may suggest that SN feedback plays a minor role in quenching, it nonetheless contributes to heating the cold gas into the hot phase, helping to set the conditions that enable AGN feedback to become effective following SMBH growth.\\
\begin{figure*}
    \centering
    \includegraphics[width=\textwidth]{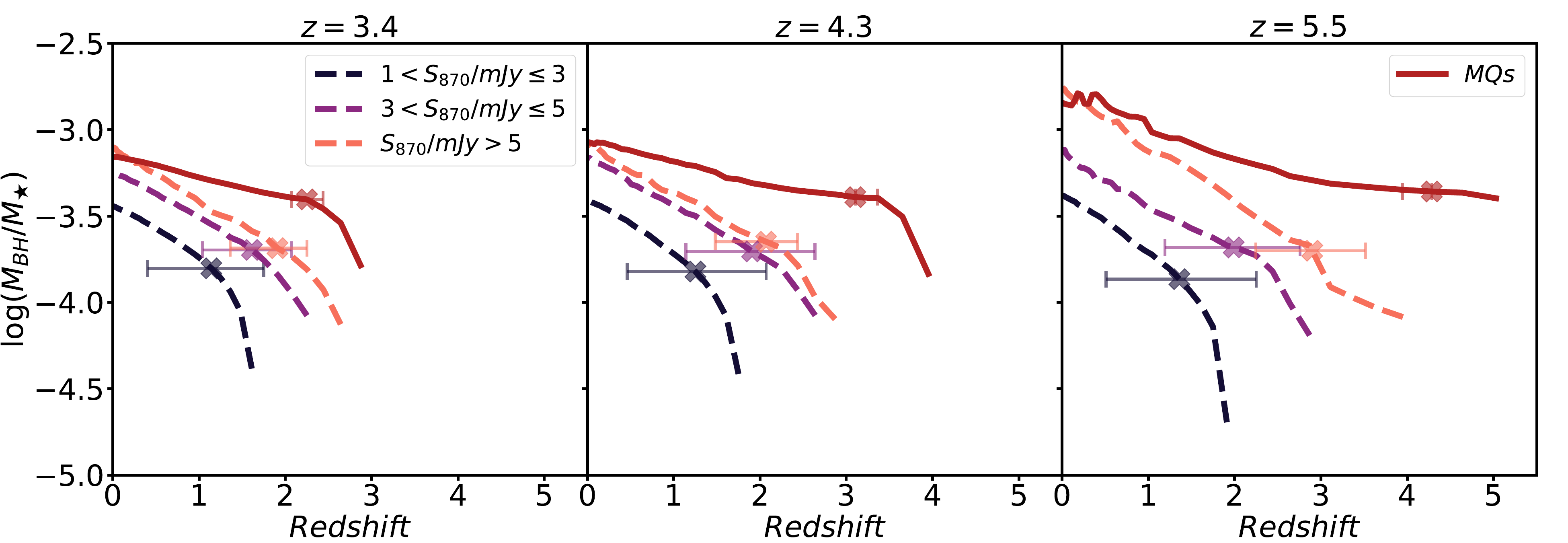}
    \caption{The evolution of the median supermassive black hole mass ($M_{\rm BH}$) to stellar mass ($M_{\star}$) ratio for MQ progenitors (red solid line), and for faint (dark blue dashed line), intermediate (purple dashed line), and bright (orange dashed line) DSFGs selected at $z=3.4$ (left-hand panel), $z=4.3$ (central panel), and $z=5.5$ (right-hand panel) — the typical redshifts at which MQs at $z = 2$, $3$, and $4$, respectively, exceeded at least $1\,{\rm mJy}$. The crosses and their error bars indicate the median, along with the $16^{\rm{th}}$ and $84^{\rm{th}}$ percentiles, of the redshift at which each galaxy population quenched. DSFGs, independently of selection redshift or sub-millimetre flux density, exhibit a similar $M_{\rm BH}/M_{\star}$ ratio at the time of quenching ($\log(M_{\rm BH}/M_{\star}) \sim -3.75$). In contrast, MQs host more massive SMBHs relative to their stellar mass at the time of quenching. This is despite the overall energy input from AGN being lower, typically (see Figure \ref{fig:feedback}), and reflects the lower stellar masses of the highest-redshift MQs at the time of quenching. In summary, in our model, rapid growth of a black hole triggered by an early galaxy-galaxy merger produces a black hole that is overmassive and sufficiently energetic to maintain the quenched state of the galaxy.}
    \label{fig:Mbh-Mstar}
\end{figure*}
\indent Interestingly, the maximum energy input rate from AGN is lower for the MQ progenitors than for the brightest classes of DSFGs, at the time at which the populations become quiescent.
This might suggest that weaker AGN feedback is required to quench star formation in MQs, but it is important to consider that the impact of this energy release on the host galaxy depends on its mass. In this context, even when presenting weaker AGN feedback, the energy input from overmassive SMBHs relative to the galaxy's gas mass may suppress the gas cooling more rapidly. To explore this, we present in Figure \ref{fig:Mbh-Mstar} the redshift evolution of the SMBH-to-stellar mass ratio ($M_{\rm BH}/M_{\star}$) for progenitors of MQs and DSFGs at $z = 3.4$, 4.3, and 5.5.\\
\indent The most remarkable trend in Figure \ref{fig:Mbh-Mstar} is that high-redshift MQ progenitors host more massive SMBHs relative to their stellar content, exhibiting higher $M_{\rm BH}/M_{\star}$ ratios than the SMBHs of typical DSFGs across cosmic time, especially at higher redshifts. In our model, SMBHs in MQs grow rapidly due to an early merger, resulting in overmassive SMBHs relative to their stellar mass. The AGN feedback associated with these SMBHs is therefore sufficiently energetic to quench the host galaxy efficiently.\\
\indent Another notable trend is that the $M_{\rm BH}/M_{\star}$ ratio correlates with the submillimetre flux density of DSFGs; bright DSFGs exhibit higher $M_{\rm BH}/M_{\star}$ ratios than faint DSFGs at any given redshift. However, at the time of quenching (indicated by crosses in Figure \ref{fig:Mbh-Mstar}), DSFGs show similar $M_{\rm BH}/M_{\star}$ ratios, with faint DSFGs having slightly lower values. Comparing the evolution of the $M_{\rm BH}/M_{\star}$ ratio across the different populations, DSFGs exhibit a more pronounced increase from high to low redshift than MQ progenitors, whose evolution proceeds at a significantly slower rate following very early black hole mass growth before quenching. This, combined with the results from Figure \ref{fig:m_channels}, which show that galaxies increase their stellar mass mainly through accretion in (dry) mergers after quenching, suggests that such events play a more important role in growing SMBHs for galaxies that quench later.\\
\begin{figure}
    \centering
    \includegraphics[width=\columnwidth]{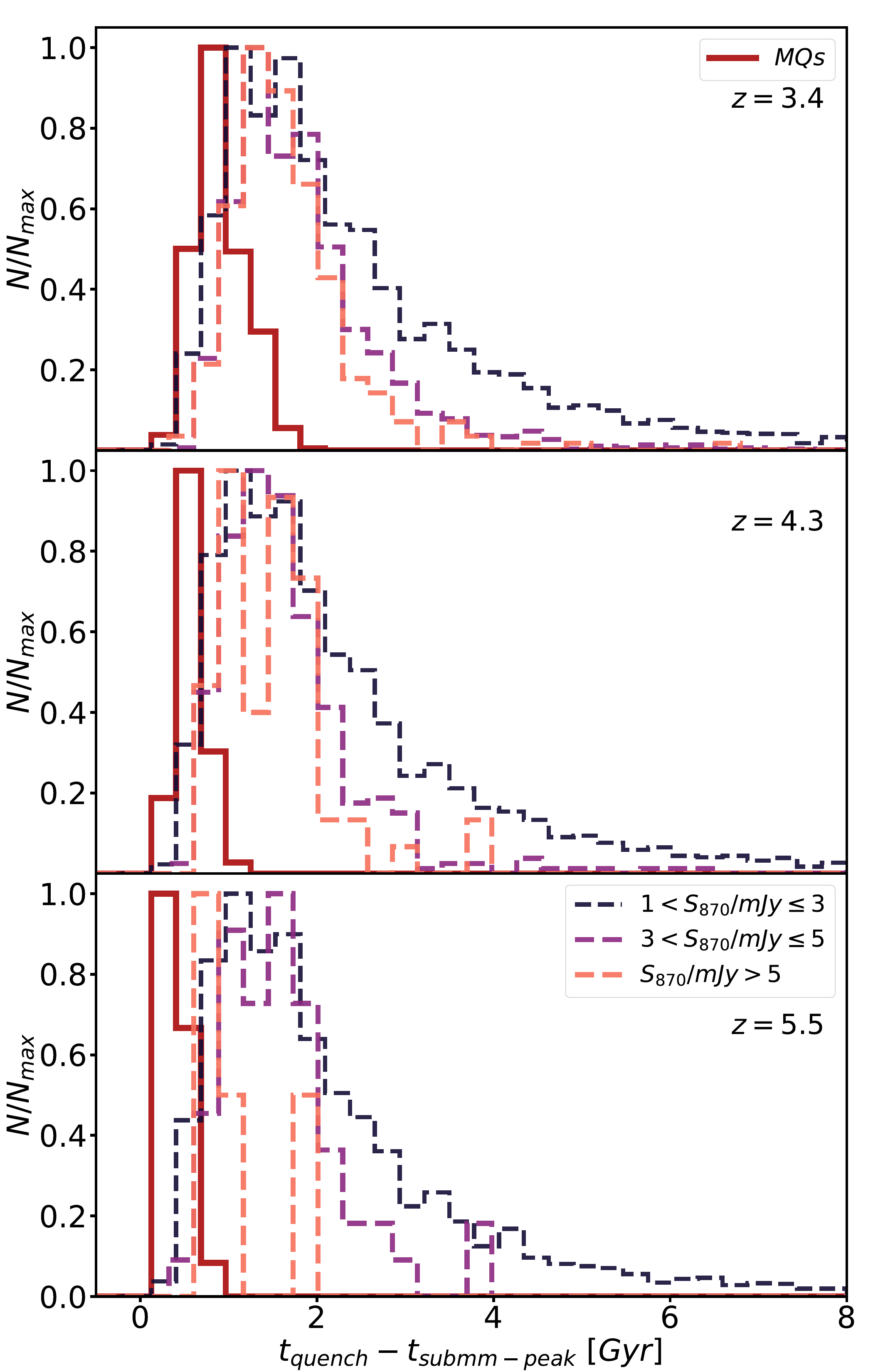}
    \caption{The distribution of timescales between the sub-millimetre peak ($S_{870, {\rm max}}$) and subsequent quenching is shown for massive quiescent (MQ) progenitors (solid red histogram), as well as for faint (dark blue dashed), intermediate (purple dashed), and bright (orange dashed) dusty star-forming galaxies (DSFGs) selected at $z = 3.4$ (first panel), $4.3$ (second panel), and $5.5$ (third panel) — the typical redshifts at which MQs at $z = 2$, $3$, and $4$, respectively, were in a DSFG phase. The timescale distributions for DSFGs exhibit similar peaks, although faint DSFGs show greater dispersion, with tails towards longer timescales. At $z = 5.5$, bright DSFGs present shorter timescales than their analogues at lower redshifts. MQ progenitors show systematically shorter timescales than DSFGs, with durations increasing for MQs selected at lower redshifts. A zoomed-in version of the MQ timescale distribution is shown in Figure~\ref{fig:DeltaTime_MQs}.}
    \label{fig:DeltaTime}
\end{figure}
\indent From Figure~\ref{fig:m_channels}, we find that, in general, all analysed galaxy populations reached their submillimetre peak close to the time of the first significant merger event. This, in turn, triggered AGN feedback (see Figure~\ref{fig:feedback}), and, in the case of high-redshift MQs, the resulting SMBH (formed via the early merger) is overmassive relative to their stellar content (see Figure~\ref{fig:Mbh-Mstar}). \\ \indent In Figure~\ref{fig:DeltaTime}, we present the timescale distributions, $t_{\rm quench} - t_{\rm submm-peak}$, for both high-redshift MQ progenitors and DSFGs selected at $z = 3.4$, 4.3, and 5.5. As anticipated from Figure~\ref{fig:Mbh-Mstar}, MQ progenitors quench rapidly after reaching their submillimetre peak due to the enhanced growth of their SMBHs relative to their stellar mass. A zoomed-in version of the MQ timescales is shown in Figure~\ref{fig:DeltaTime_MQs}. On average, MQs observed at $z = 2$, 3, and 4 quenched their star formation approximately $0.9$~Gyr, $0.6$~Gyr, and $0.4$~Gyr, respectively, after reaching their submillimetre peak. This indicates that MQs selected at higher redshift generally quenched more rapidly than those at lower redshift. In contrast, the timescales between the submillimetre peak and quenching for DSFGs are longer and exhibit greater dispersion, particularly for the faintest sources. For instance, although the peak of the distributions is similar among the DSFG populations, the median timescales for faint DSFGs at $z = 3.4$, 4.3, and 5.5 are consistently around $\sim 1.95 \pm 0.1$~Gyr. Similarly, the median timescales for intermediate DSFGs do not vary significantly across the redshifts analysed, with $t_{\rm quench} - t_{\rm submm-peak} \sim 1.32$--$1.46$~Gyr. In contrast, the timescales for bright DSFGs selected at $z = 3.4$, 4.3, and 5.5 are $\sim 1.37$~Gyr, $1.32$~Gyr, and $0.8$~Gyr, respectively. This trend highlights the different evolutionary pathways followed by DSFGs, which we will explore further in a forthcoming study.

\section{Discussion} \label{sec:discussion}
Observational results suggest a connection between dusty star-forming galaxies (DSFGs) and massive quiescent galaxies (MQs) at high-redshift, supported by the typical inferred star-formation histories of MQs (i.e. intense, short-lived starbursts; e.g. \citealt{glazebrook17, forrest20, carnall23b}), number densities \citep{valentino20}, and their host halo masses, e.g. \citealt[][]{Wilkinson17}). However, this potential link — where DSFGs serve as progenitors of high-redshift MQs — has not yet been thoroughly explored within the galaxy formation modelling framework, as simulations often struggle to reproduce DSFG and MQ populations simultaneously. Here, we used one of the semi-analytic models from \citet{yo25a}, which provides a reasonable representation of these populations, to investigate the connection between DSFGs and MQs.

\subsection{The connection between dusty star-forming and massive quiescent galaxies at high-redshift} \label{sec:disc-connection}

The model from \citet{yo25a} predicts that approximately $86\%$, $96\%$, and $91\%$ of MQs at $z = 2$, 3, and 4, respectively, exceeded a flux density of $1\,\mathrm{mJy}$ in the $870\,\mu$m band at some point in their past ($S_{870,\mathrm{max}}$; see Figure~\ref{fig:S870-dist}), indicating a strong link between DSFGs and MQs at high redshift. However, only a small fraction ever reached $S_{870} \geq 3\,\mathrm{mJy}$ (around $7\%$, $9\%$, and $5\%$ of MQs at $z = 2$, 3, and 4, respectively), suggesting that while most MQs were once DSFGs, more extreme DSFGs have a higher probability of quenching earlier and becoming MQs at $z > 2$. This is broadly consistent with the findings of \citet{valentino20}, who also examined this connection using \texttt{IllustrisTNG}. However, our model indicates a stronger relationship, with nearly all MQs at high redshift reaching at least $1\,\mathrm{mJy}$. This discrepancy may arise because the \texttt{IllustrisTNG} model underestimates submillimetre number counts \citep[see][]{chris21}.\\
\indent We derive and analyse the redshift distribution at which MQs reach their maximum submillimetre emission (Figure~\ref{fig:S870-dist}, bottom panels). For MQs selected at $z = 2$, the peak submillimetre flux density occurs at $z(S_{870,\mathrm{max}}) = 2.9-4.0$ ($16^{\rm{th}}-84^{\rm{th}}$ percentile), with a median of $z = 3.2$. Similarly, for MQs selected at $z = 3$, the distribution spans $z = 4.0-5.0$, peaking at $z = 4.3$, while MQs at $z = 4$ show $z(S_{870,\mathrm{max}}) = 5.0-5.9$ and a median of $z = 5.5$. This trend is expected, as some MQs at a given redshift could have quenched earlier, and there is less cosmic time before higher-redshift snapshots; indeed, we find that MQs exhibit similar quenching timescales after reaching their submillimetre peak (Figure~\ref{fig:DeltaTime}).\\
\indent Although the redshift ranges of peak emission are similar, this does not imply that MQ progenitors become DSFGs only at the time of $S_{870,\mathrm{max}}$. As shown in Figure~\ref{fig:S870-track_MQs}, the submillimetre flux density of an individual galaxy increases gradually with decreasing redshift towards its peak. In Figure~\ref{fig:Fraction-S870-z}, we quantify the fraction of MQ progenitors that exceed a given $S_{870}$ threshold as a function of redshift. We find that $50\%$, $66\%$, and $67\%$ of MQs selected at $z = 2$, 3, and 4, respectively, surpass $S_{870} = 1\,\mathrm{mJy}$ at $z \sim 3.4$, 4.3, and 5.5. Based on these values, we select the full DSFG population at these corresponding redshifts.\\
\indent We find that a large fraction of DSFGs at $z \gtrsim 3$ do not evolve into MQs by $z \gtrsim 2$. In Figure~\ref{fig:Fraction-SMG-Q}, we quantify the fraction of DSFGs selected at $z = 3.4$, $4.3$, and $5.5$ that subsequently evolve into massive and quiescent systems by a given redshift. While the typical MQ progenitors reach $S_{870} \sim 1-3\,\mathrm{mJy}$, less than $10\%$ of the total faint DSFG population ($S_{870} \lesssim 3\,\mathrm{mJy}$) quench by the selected epochs. This result implies that the majority of DSFGs do not follow the same evolutionary pathway as MQ progenitors (see Section~\ref{sec:disc-phys}).\\
\indent Further evidence supporting the DSFG–MQ connection is presented in Figure~\ref{fig:S870-smass}, which shows the relationship between the stellar mass of MQs and their historical maximum $S_{870}$ flux density. The clear, nearly redshift-independent correlation between stellar mass (an observed-day property) and $S_{870,\mathrm{max}}$ (a proxy for past star formation) implies that the most massive MQs were also the brightest in the submillimetre regime. This suggests that extreme DSFGs ($S_{870} \gtrsim 5\,\mathrm{mJy}$) may indeed be the progenitors of the most massive ($M_\star \gtrsim 10^{11}\,\rm{M_\odot}$) quiescent galaxies at high-redshift. This relation may reflect the early evolution of massive systems, as the overall DSFG population follows the $M_{\star}$–SFR main sequence (see Figure \ref{fig:SFR-Mstar}). In this framework, higher stellar masses correspond to higher SFRs, which lead to enhanced dust production (from supernovae) and consequently to higher $S_{870}$ flux densities. 
\subsection{How dusty star-forming galaxies become massive quiescent at high-redshift} \label{sec:disc-phys}
From our model, we find that not all DSFGs evolve rapidly into massive quiescent galaxies. We investigate the physical mechanisms that transform a subset of DSFGs into MQs. First, we analyse the median stellar mass growth through secular processes, accretion via mergers, and merger-induced starbursts for DSFGs with different submillimetre flux densities, along with MQ progenitors (Figure~\ref{fig:m_channels}). The key difference in stellar mass assembly between these populations lies in the timing of their first major merger. In the \texttt{L-Galaxies} semi-analytic model, mergers trigger starbursts, and in the \citet{yo25a} framework, this mechanism converts the available cold gas into stars with high efficiency. For example, in equal-mass mergers, approximately $90\%$ of the cold gas is consumed to form new stars. As a result, a significant fraction of the final stellar mass — compared to the previously in-situ-formed mass — originates from merger-induced starbursts in MQs selected at a given redshift. We also note that the starburst event typically coincides with the epoch at which MQs reach their maximum submillimetre flux density. This is consistent with the findings of \citet{forrest20}, who observed that ultramassive galaxies at $z > 3$ that have already quenched display post-starburst signatures.

As noted above, an early major merger dictates the evolutionary paths followed by the general DSFG population and high-redshift MQ progenitors. Besides inducing starbursts, mergers are also the main driver of supermassive black hole growth in \texttt{L-Galaxies}. In this context, merger events can simultaneously boost both supernova (SN) and AGN feedback. To identify the dominant mechanism responsible for quenching high-redshift MQs, we present in Figure~\ref{fig:feedback} the median energy injection from SN and AGN feedback as a function of redshift for the populations under study. As expected, AGN feedback dominates, with energy input exceeding that of SN feedback by at least two orders of magnitude near the quenching epoch. This scenario is consistent with observational results: massive galaxies at $z \gtrsim 3$ often host AGN \citep{marchesini10, stefanon15, marsan17, forrest20}, which become more luminous (a proxy for AGN feedback strength) with increasing host stellar mass, particularly in already quenched systems \citep{ito22}.

Figure~\ref{fig:feedback} thus supports a scenario in which both DSFGs and high-redshift MQs quench primarily due to AGN feedback, triggered by mergers that are also associated with the submillimetre peak. However, as shown in Figure~\ref{fig:DeltaTime}, high-redshift MQ progenitors quench more rapidly than the overall DSFG population after reaching their submillimetre peak. We explore this in Figure~\ref{fig:Mbh-Mstar}, where we present the evolution of the SMBH-to-stellar mass ratio for the various populations. We find that the rapid quenching of high-redshift MQs is driven by early mergers that result in overmassive SMBHs relative to the stellar mass of MQ progenitors. Consequently, less AGN feedback energy is required to quench star formation in these systems. 

Our predicted scenario is broadly consistent with the results of \citet{Xie24} based on the \texttt{GAEA} SAM, where mergers trigger AGN feedback that ultimately quenches star formation. Our prediction of overmassive black holes in MQs, relative to their stellar mass when compared to star-forming galaxies of similar stellar mass (in our case, DSFGs), is also reproduced in the \texttt{IllustrisTNG} and \texttt{ASTRID} simulations \citep{Weller25}. However, in contrast to our model, mergers in \texttt{IllustrisTNG} do not play a significant role in suppressing star formation in MQs at $z \gtrsim 3$, as shown by \citet{Kurinchi-Vendhan24}; this is likely due to the different model for black hole growth in \texttt{IllustrisTNG}.

\subsection{Limitations and caveats} \label{sec:disc-lims}
As discussed in \citet{yo25a}, the galaxy formation model used in this work (referred to as `no HIMF') is a re-calibrated version of the \citet{henriques20} version of \texttt{L-Galaxies}, and, like other eight recalibrated variants, it still presents certain limitations. 

Similarly to the other eight alternative versions obtained in \citet{yo25a}, the model adopted here struggles to reproduce the high-mass end of the stellar-mass function at high redshift. Additionally, it overpredicts the cosmic star-formation rate density by $\sim0.5\,\rm{dex}$ relative to observational constraints. However, the `no HIMF' model (used in this work) is among the best-performing at reproducing the number density of MQs at high redshift, showing consistency with the lower limits from observations. Nevertheless, it still underpredicts the number density by approximately $1\,\rm{dex}$ compared to recent JWST measurements at $z \sim 3.5$. In contrast, six out of the nine models match or closely match the observed submillimetre number counts; the model used here is not the best of these models — albeit it is still in good agreement. As noted in Section~\ref{sec:disc-connection}, this may result in a failure to capture MQs that do not descend from DSFGs.

One likely cause of the underprediction of the high-redshift MQ population is the simplified AGN-feedback and black hole growth prescription adopted in the \citet{henriques20} version of \texttt{L-Galaxies}, originally introduced by \citet{croton06}. In the \citet{croton06} framework, SMBH growth occurs mainly through mergers, which can lead to an underestimation of SMBH masses in the early Universe ($z \gtrsim 4$). In the context of the `no HIMF' recalibrated version presented in \citet{yo25a}, a $\lesssim 3\%$ fraction (scaled by the mass ratio) of the cold gas is accreted during mergers, independently of the halo mass. Among all recalibrated variants, the `no HIMF’ model is, in fact, one of the most efficient at accreting cold gas onto SMBHs. However, because this efficiency is independent of halo mass, the resulting $M_{\rm BH}-M_{\star}$ relation remains almost constant at high redshift (see Figure~\ref{fig:Mbh-Mstar-all}), leading to an underprediction of SMBH masses in massive galaxies compared with other models \citep[see Figure 13][]{lagos25} and observational data from \citet{Decarli10a, Decarli10b}, \citet{Suh20}, and \citet{Poitevineau23} at $z \sim 3$. A more sophisticated treatment of SMBH growth and formation, as implemented in the \citet{izquierdo22} and \citet{spinoso23} versions of \texttt{L-Galaxies}, improves consistency with current observations. Nevertheless, our model performs reasonably well in reproducing the observed SMBH mass function and the $M_{\rm BH}-M_{\star}$ relation at low redshift. 

As shown in this study, AGN feedback is the primary mechanism responsible for quenching star formation in both DSFGs and MQs at high redshift. In the \citet{croton06} model, AGN feedback operates exclusively on the hot gas halo, suppressing cooling flows without directly affecting the cold gas reservoir of the galaxy. In the `no HIMF' recalibrated model, AGN feedback is stronger than in the default \citet{henriques20} version. However, it still underpredicts the fraction of quiescent galaxies at high redshift—likely by the underestimation of SMBH masses at early times, as discussed above. As shown by \cite{lagos24} and \citet{delucia24} for the \texttt{SHARK} and \texttt{GAEA}\footnote{In fact, the SMBH–related processes prescription adopted in \texttt{GAEA} builds on the \citet{croton06} framework \citep{Fontanot20}, but additionally includes cold gas accretion as a channel for SMBH growth, as well as AGN-driven winds that directly affect the cold gas in galaxy disks.}  SAMs, implementing a more comprehensive model for AGN-driven feedback, capable, for example, of heating or ejecting cold gas, may help resolve the strong discrepancy with observed MQ abundances. Increasing the number density of MQs at high redshift would naturally enhance the predicted transition rate between DSFGs and MQs.

A key finding from \citet{yo25a} is the high degree of parameter degeneracy in \texttt{L-Galaxies}, likely in all galaxy formation models, due to their large parameter space (15 free parameters in the case of \texttt{L-Galaxies}). Thus, although our model provides a plausible evolutionary pathway for high-redshift MQs that broadly agrees with observations, alternative models featuring different physical assumptions may also be capable of reproducing both the DSFG and MQ populations.

\subsection{The nature of dusty star-forming galaxies}
In this work, we explored the connection between DSFGs and MQs at high redshift, with a particular focus on the progenitors of MQs. We have shown that the progenitors of $z > 2$ MQs largely follow similar evolutionary pathways to faint DSFGs, characterised by submillimetre flux densities in the range $1 \, {\rm mJy} < S_{870} \leq 3 \, {\rm mJy}$ at early times ($z \gtrsim 5$).

In comparing with the broader DSFG population, we found that these galaxies typically exceed the $S_{870} > 1 \, {\rm mJy}$ threshold at $5 \lesssim z \lesssim 7$, depending on the redshift at which they are selected. DSFGs become submillimetre-bright primarily due to secular processes (Figure~\ref{fig:m_channels}), and reach their submillimetre flux peak, on average, around the time of their first significant merger event—which occurs later than in the case of high-redshift MQ progenitors. Moreover, unlike the high-redshift MQ progenitors, the quenching timescales after the submillimetre peak in DSFGs span a wide range, from a few hundred Myr up to $\sim4\,\mathrm{Gyr}$ in the case of the brightest DSFGs.

A natural extension of this study will be to characterise the different evolutionary pathways followed by DSFGs, along with their progenitors and descendants. As discussed in the previous subsection, \citet{yo25a} identified five additional models with different physical prescriptions (different free parameters) that yield submillimetre number counts consistent with observations. In future work, we will analyse the formation and evolution of DSFGs across these alternative models, in order to assess the robustness of our conclusions.

\section{Summary} \label{sec:summary}
In this work, we present theoretical predictions regarding the potential connection between dusty star-forming galaxies (DSFGs) and massive quiescent galaxies (MQs) at high redshift ($z = 2$, 3, and 4). We use the \citet{henriques20} version of the \texttt{L-Galaxies} SAM, specifically employing the set of free parameters obtained in \citet{yo25a}, which succeeds in reproducing both galaxy populations studied in this work.
Our main findings are summarised as follows: 

\begin{enumerate} 
\item A total of $86\%$, $96\%$, and $90\%$ of MQs at $z = 2$, $3$, and $4$, reached at least $1\,\rm{mJy}$ in sub-millimetre flux density at some point in their past. Hence, the vast majority of MQs are the descendants of $z \gtrsim 3.4$ DSFGs (see Figure \ref{fig:S870-dist}, top panels). The typical redshifts at which these high-redshift MQs reached their maximum sub-millimetre emission ($S_{870,\mathrm{max}}$) are $z \sim 3.1$, $4.3$, and $5.5$, for the MQs at $z = 2$, $3$, and $4$, respectively (Figure \ref{fig:S870-dist}, bottom panels).

\item We find a strong correlation between the stellar mass of high-redshift MQs and the maximum sub-millimetre flux density they achieved in their past, $S_{870,\mathrm{max}}$ (Figure \ref{fig:S870-smass}). The most massive MQs had the brightest past sub-millimetre emission. A linear fit yields a nearly redshift-independent relation, suggesting the derived functions hold for MQs in the range $2 \lesssim z \lesssim 4$.

\item Since high-redshift MQ progenitors can exhibit significant sub-millimetre emission at redshifts different from when they reach their peak, we show in Figure \ref{tab:S870-smass} the fraction of MQs that exceeded given $S_{870}$ thresholds as a function of redshift. Sub-millimetre emission was most prevalent at $z \sim 3.4$, $4.3$, and $5.5$ for MQs selected at $z = 2$, $3$, and $4$, respectively. At these epochs, approximately $50\%$, $66\%$, and $67\%$ of the MQs at the previously mentioned redshifts, exceeded $1\,\rm{mJy}$.

\item Although $z \sim 3.4$, 4.3, and $5.5$ were the redshifts when most MQ progenitors were sub-millimetre bright, only a small fraction of the broader populations of DSFGs at those redshifts are progenitors of MQs at $z = 2$, $3$, and $4$ (Figure \ref{fig:Fraction-SMG-Q}). In fact, while the brightest DSFGS begin to quench earlier, on average, the majority of DSFGs tend to quench at $z \lesssim 2$.

\item The evolution of stellar mass growth through secular star formation, merger accretion, and merger-induced starbursts (Figure \ref{fig:m_channels}) indicates that while sub-millimetre emission is mostly driven by secular star formation, the highest $S_{870}$ values are more closely associated with mergers and starbursts.

\item MQ progenitors typically experience a major merger earlier than normal DSFGs (see Figure \ref{fig:m_channels}). Within the \texttt{L-Galaxies} framework, this early merger is responsible for depleting the cold gas reservoir -- by starburst and cold gas accretion of the supermassive black hole --- and initiating the quenching process -- mostly driven by AGN feedback.

\item Our model predicts that, following the extreme merger event, these galaxies predominantly grow through merger accretion without significant starbursts, i.e. through dry mergers.

\item Finally, we trace the median evolution of energy injection from supernovae and AGN feedback for the studied galaxy populations (Figure \ref{fig:feedback}). Our results indicate that AGN feedback is the primary quenching mechanism in all cases. This AGN feedback is triggered earlier in high-redshift MQ progenitors due to mergers. While SN feedback plays a secondary role, it is essential for heating cold gas and preparing the conditions necessary for AGN activation. 
\end{enumerate}

These results offer predictions and a detailed theoretical framework for understanding the progenitors of massive quiescent galaxies at high redshift and their connection to dusty star-forming galaxies. We find that high-redshift MQs were, in most cases, previously DSFGs. However, most DSFGs only quench at $z \lesssim 2$. The key distinction between high-redshift MQ progenitors and the broader DSFG population lies in the timing of the first significant merger event, which triggers AGN activity and ultimately quenches star formation.
\begin{acknowledgements}
      We thank the anonymous referee for constructive comments that helped improve the clarity of this work.
      PA-A thanks the Coordenaç\~ao de Aperfeiçoamento de Pessoal de Nível Superior – Brasil (CAPES), for supporting his PhD scholarship (project 88882.332909/2020-01). RKC is grateful for support from the Leverhulme Trust via the Leverhulme Early Career Fellowship. The model calibration was run on computing cluster facilities provided by the Flatiron Institute. The Flatiron Institute is supported by the Simons Foundation. LSJ acknowledges the support from CNPq (308994/2021-3)  and FAPESP (2011/51680-6). MCV acknowledges the FAPESP for supporting his PhD (2021/06590-0). PA-A and BG acknowledge support from the Independent Research Fund Denmark (DFF) under grant 4251-00086B. FV acknowledges support from the DFF under grant 3120-00043B.
\end{acknowledgements}

\bibliographystyle{bibtex/aa.bst}
\bibliography{sample631.bib}

\begin{appendix}
\section{Energy input due to feedback} \label{sec:energy-inputs}
The energy inputs due to SN and AGN feedback, $\Delta E_{\rm SN}$ and $\Delta E_{\rm AGN}$, (see Figure \ref{fig:feedback}), are not direct outputs from \texttt{L-Galaxies}; however, they can be easily derived from other properties. 
To calculate the energy input from SN feedback, we use the outputs \texttt{ReheatingRate} ($\Delta M_{\rm reheat}$) and \texttt{EjectionRate} ($\Delta M_{\rm eject}$), which correspond to the mass rates of reheated and ejected gas, respectively. It is important to note that, in \texttt{L-Galaxies}, ejection occurs only if there is remaining energy after reheating. The total energy input is then given by Equation~\ref{eq:feedback-SN} — Adapted from Equation~S23 in the Supplementary Material of \citet{henriques20}:
\begin{equation}
\Delta E_{\rm SN} = \frac{1}{2} (\Delta M_{\rm reheat} + \Delta M_{\rm eject}) \, V_{200c}^2 \ ,
\label{eq:feedback-SN}
\end{equation}
where $V_{200c}$ is the halo virial velocity.

The AGN feedback in the \citet{henriques20} version of \texttt{L-Galaxies} is based on the \citet{croton06} model. To obtain the energy input due to AGN feedback, we first estimate the accretion using Equation \ref{eq:feedback-AGN}:
\begin{equation}
\Delta M_{\rm BH} = k_{\rm AGN} \left( \frac{M_{\rm hot}}{10^{11} \rm{M_{\odot}}} \right) \left( \frac{M_{\rm BH}}{10^{8} \rm{M_{\odot}}} \right) \frac{H_0}{H(z)} \ ,
\label{eq:feedback-AGN}
\end{equation}
where $k_{\rm AGN}$ is a free parameter ($k_{\rm AGN} = 0.07\,\rm{M_{\odot}}{\rm yr}^{-1}$ in the recalibrated model used in this work), $M_{\rm hot}$ is the mass of hot gas, and $M_{\rm BH}$ is the supermassive black hole mass. Finally, the energy input due to AGN feedback is given by $\Delta E_{\rm AGN} = \eta \, \Delta M_{\rm BH} \, c^2$ (Equation~S29 in the Supplementary Material of \citet{henriques20}), where $\eta = 0.1$ and $c$ is the speed of light. 

\section{Timescales between the peak of sub-millimetre emission and subsequent quenching}

\begin{figure}
    \centering
    \includegraphics[width=\columnwidth]{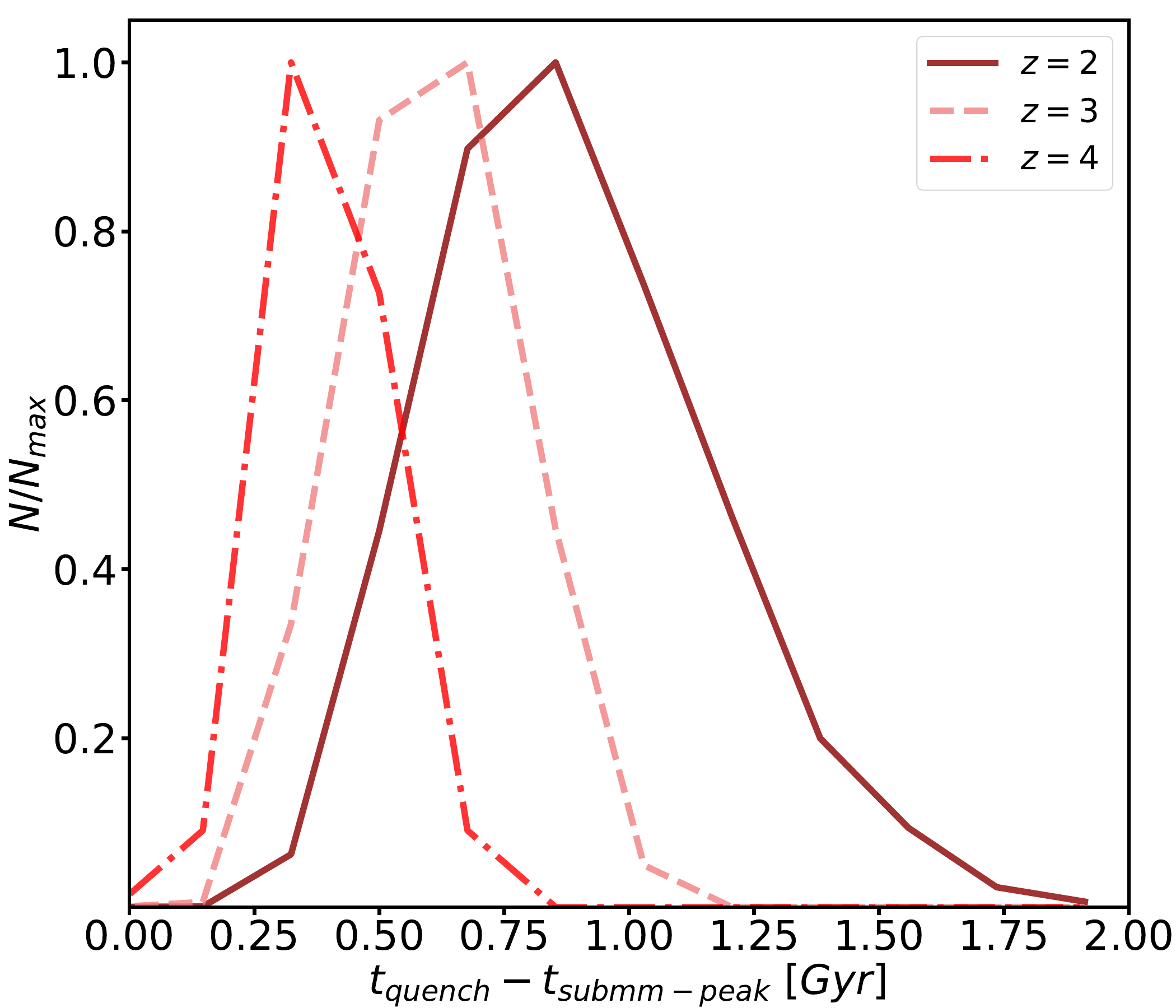}
    \caption{A zoomed-in version of Figure \ref{fig:DeltaTime}, showing only massive quenched galaxies at $z=2$, $3$ and $4$. The time between the peak in sub-millimetre flux density and subsequent quenching is lowest for the highest redshift MQs, on average. The median $t_{\rm quench} - t_{\rm submm-peak}$ timescales are $0.9\,$Gyr, $0.6\,$Gyr, and $0.4\,$Gyr, for MQs selected at $z = 2$, $3$, and $4$, respectively.}
\label{fig:DeltaTime_MQs}
\end{figure}

Figure \ref{fig:DeltaTime} shows the distribution of timescales between the sub-mm emission peak ($t_{\rm submm-peak}$) and subsequent quenching ($t_{\rm quench}$) for DSFGs and MQ progenitors at $z = 3.4$, 4.3, and 5.5. For visualization purposes, Figure \ref{fig:DeltaTime_MQs} presents a zoom-in view of Figure \ref{fig:DeltaTime}, focusing exclusively on MQs selected at $z = 2$, 3, and 4. The main trend observed in Figure \ref{fig:DeltaTime_MQs} is that, on average, MQs at higher redshifts quench more rapidly after reaching their sub-millimetre peak than those identified at lower redshifts.

We find that the median $t_{\rm quench} - t_{\rm submm-peak}$ timescales are $0.9\,$Gyr, $0.6\,$Gyr, and $0.4\,$Gyr, for MQs selected at $z = 2$, $3$, and $4$, respectively. 

\section{$M_{\rm BH}$-$M_{\star}$ Relation}
\begin{figure}
    \centering
    \includegraphics[width=\columnwidth]{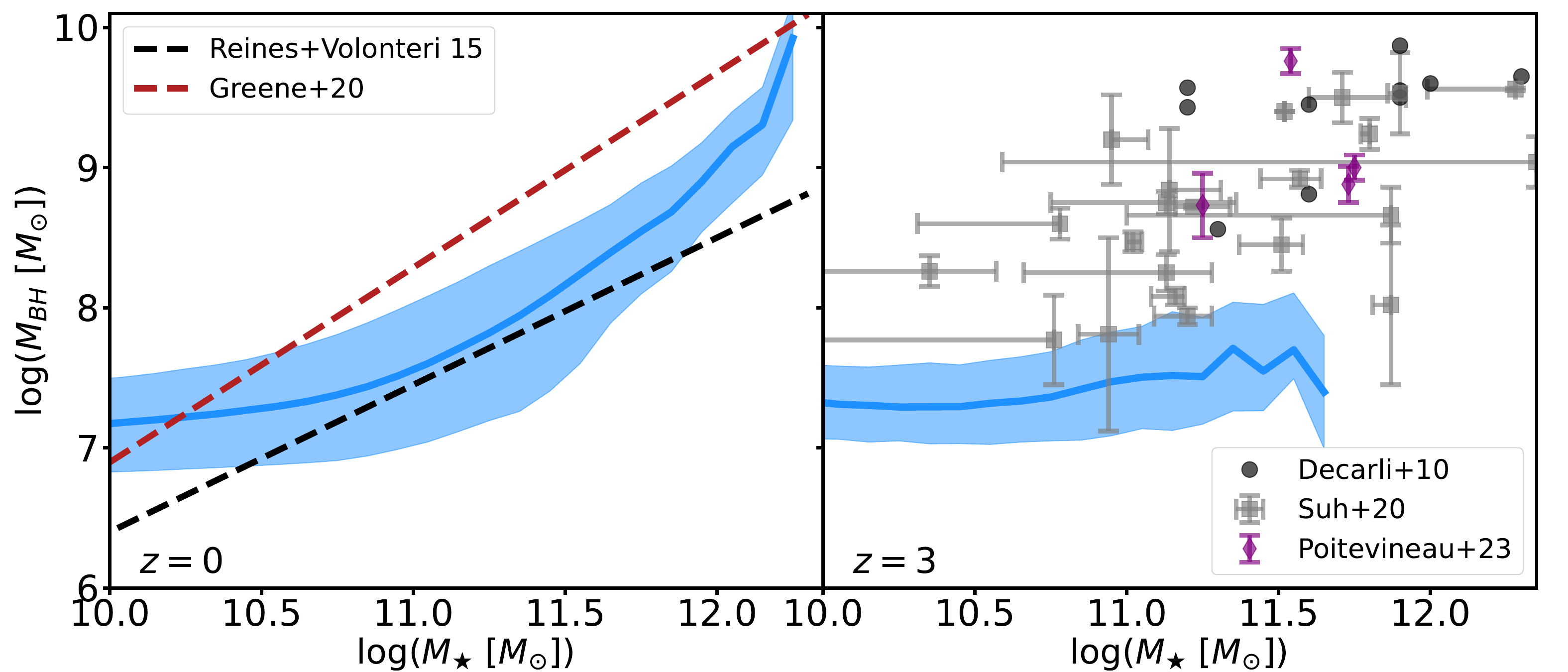}
    \caption{The $M_{\rm BH}-M_{\star}$ relation at $z = 0$ (left) and 3 (right) as predicted by the `no HIMF' recalibrated model \citep{yo25a}. The light-blue areas and lines represent the $16^{\rm th}$, $50^{\rm th}$, and $84^{\rm th}$ percentiles of all galaxies in the simulated box. The black and red dashed lines indicates the $M_{\rm BH}-M_{\star}$ relations derived from observations by \citet{Reines15} and \citet{Greene20}, respectively. The predictions at $z = 3$ are compared to observational estimates from \citet{Decarli10a, Decarli10b} (black dots), \citet{Suh20} (gray squares) and \citet{Poitevineau23} (purple diamonds).}
\label{fig:Mbh-Mstar-all}
\end{figure}

For completeness and as a sanity check, we present in Figure~\ref{fig:Mbh-Mstar-all} the $M_{\rm BH}$–$M_{\star}$ relation for the `no HIMF’ model of \citet{yo25a} at $z = 0$ and $z = 3$. At $z = 0$, we compare our predictions with the observational relations from \citet{Reines15} and \citet{Greene20}. The observational estimates from \citet{Decarli10a, Decarli10b, Suh20} and \citet{Poitevineau23} at $2.0 < z < 3.5 $ are compared against our predictions at $z = 3$. 

\end{appendix}
\end{document}